\documentclass[twocolumn,trackchanges]{aastex631} 
\usepackage{amsmath}
\usepackage{multirow}

\newcommand{\mytilde}{\raise.19ex\hbox{$\scriptstyle\sim$}}

\shorttitle{NIRWL PSF and systematics}
\shortauthors{Finner et al.}
\graphicspath{{./}{figures/}}

\begin{document}

\title{Near-IR Weak-Lensing (NIRWL) Measurements in the CANDELS Fields I: Point-Spread Function Modeling and Systematics}

\correspondingauthor{Kyle Finner}
\email{kfinner@caltech.edu}

\author[0000-0002-4462-0709]{Kyle Finner}
\affiliation{IPAC, California Institute of Technology, 1200 E California Blvd., Pasadena, CA 91125, USA}

\author[0000-0003-1954-5046]{Bomee Lee}
\affiliation{Korea Astronomy and Space Science Institute (KASI), 776 Daedeokdae-ro, Yuseong-gu, Daejeon 34055, Republic of Korea}

\author[0000-0001-7583-0621]{Ranga-Ram Chary}
\affiliation{IPAC, California Institute of Technology, 1200 E California Blvd., Pasadena, CA 91125, USA}

\author[0000-0002-5751-3697]{M. James Jee}
\affiliation{Department of Astronomy, Yonsei University, 50 Yonsei-ro, Seodaemun-gu, Seoul 03722, Republic of Korea}
\affiliation{Department of Physics and Astronomy, University of California, Davis, One Shields Avenue, Davis, CA 95616, USA}

\author[0000-0002-2951-4932]{Christopher Hirata}
\affiliation{Center for Cosmology and Astroparticle Physics, The Ohio State University, 191 West Woodruff Ave, Columbus, Ohio 43210, USA}
\affiliation{Department of Physics, The Ohio State University, 191 West Woodruff Ave, Columbus, Ohio 43210, USA}
\affiliation{Department of Astronomy, The Ohio State University, 140 West 18th Ave, Columbus, Ohio 43210, USA}

\author[0000-0003-2508-0046]{Giuseppe Congedo}
\affiliation{ Institute for Astronomy, University of Edinburgh, Royal Observatory, Blackford Hill, Edinburgh, EH9 3HJ, United Kingdom}

\author[0000-0000-0000-0000]{Peter Taylor} 
\affiliation{Center for Cosmology and Astroparticle Physics, The Ohio State University, 191 West Woodruff Ave, Columbus, Ohio 43210, USA}
\affiliation{Department of Physics, The Ohio State University, 191 West Woodruff Ave, Columbus, Ohio 43210, USA}
\affiliation{Department of Astronomy, The Ohio State University, 140 West 18th Ave, Columbus, Ohio 43210, USA}

\author[0000-0002-2550-5545]{Kim HyeongHan} 
\affiliation{Department of Astronomy, Yonsei University, 50 Yonsei-ro, Seodaemun-gu, Seoul 03722, Republic of Korea}



\begin{abstract}

We have undertaken a Near-IR Weak Lensing (NIRWL) analysis of the CANDELS HST/WFC3-IR F160W observations. With the Gaia proper-motion-corrected catalog as an astrometric reference, we updated the astrometry of the five CANDELS mosaics and achieved an absolute alignment within $0.02\pm0.02\arcsec$ on average, which is a factor of several superior to existing mosaics. These mosaics are available to download\footnote{\href{https://archive.stsci.edu/hlsp/candelsnirwl}{Link to download NIRWL mosaics.}}. We investigated the systematic effects that need to be corrected for weak-lensing measurements. We find the largest contributing systematic effect is caused by undersampling. We find a sub-pixel centroid dependence on the PSF shape that causes the PSF ellipticity and size to vary by up to 0.02 and $3\%$, respectively. Using the UDS as an example field, we show that undersampling induces a multiplicative shear bias of -0.025. We find that the brighter-fatter effect causes a 2\% increase in the size of the PSF and discover a brighter-rounder effect that changes the ellipticity by 0.006. Based on the small range of slopes in a galaxy's spectral energy distribution (SED) within the WFC3-IR bandpasses, we suggest that the impact of galaxy SED on PSF is minor. Finally, we modeled the PSF of WFC3-IR F160W for weak lensing using a principal component analysis. The PSF models account for temporal and spatial variations of the PSF. The PSF corrections result in residual ellipticities and sizes, $|de_1| < 0.0005\pm0.0003$, $|de_2| < 0.0005\pm0.0003$, and $|dR| < 0.0005\pm0.0001$, that are sufficient for the upcoming NIRWL search for massive overdensities in the five CANDELS fields.

\end{abstract}

\keywords{HST, Euclid, Roman, PSF, WL}


\section{Introduction} \label{sec:intro}
Weak gravitational lensing is a powerful tool for studying the distribution of mass in the universe. Enormous efforts have been made in understanding the large-scale structure of the universe by mapping it with weak lensing. Targeted studies focus on the mapping of dark matter in galaxy clusters, the most massive objects in the universe \citep[large survey examples are][]{2011jee, 2014umetsu, 2018bschrabback}. 
In this regime, the weak-lensing signal is abundant.  For measuring weak-lensing among lower mass groups or from galaxy-galaxy lensing, one needs to focus on the stacked lensing signal \cite[for example][]{2012lin, 2013heymans, 2016jee_shear, 2021asgari, 2022secco}. Each of these probes of the universe have been fruitful in advancing our knowledge in cosmology and have triggered new weak-lensing surveys with future telescopes such as the Vera Rubin Observatory \citep{2019ivezi}, Euclid \citep{2011laureijs, 2022scaramella}, and the Nancy Grace Roman Telescope \citep{2015spergel, 2019akeson}. These surveys plan to cover a substantial amount of the sky with the former two performing weak-lensing measurements 
in the optical regime and the latter in the infrared (IR).

Weak lensing is a statistical analysis of galaxy shapes. 
Therefore, it is desirable to maximize the number of galaxies detected while simultaneously performing accurate shape measurements. Near-IR (NIR) light offers significant advantages over optical light in some aspects of weak lensing.
\cite{2018lee} showed that NIR light provides a 1.4 factor increase in galaxy number density over optical light for the $z>1$ universe. The gain in detection is driven by the higher surface brightness of galaxies in the NIR, which probes the older stellar population that is less affected by dust. 
In typical weak-lensing analyses, the $z>1$ galaxies are a major contributor to the lensing signal. Furthermore, the older stellar population of the galaxy has a smoother light profile, which allows a lower shape noise to be achieved \citep{2018schrabback}. NIR detectors offer advantages for weak lensing as well because they do not suffer from charge transfer inefficiency (CTI), which causes significant bleeding of charge along the readout direction of optical CCDs and leads to biases in shape measurements \citep{2010rhodes}. Even though the benefits of NIR weak lensing are clear, very few studies have utilized the power. 
But with the upcoming wide surveys of Euclid and Roman probing the universe in NIR bands, it is an opportune time to demonstrate the power of NIR weak lensing and to investigate the systematic effects that come with the territory. 


With that in mind, we have initiated a weak-lensing study of the five CANDELS \citep{2011grogin,2011koekemoer} fields to examine the difficulties of weak-lensing analysis in the NIR. In this paper, we investigate the systematics of PSF modeling and shape measurement in the NIR. Bringing the relative significance of the different systematics to the attention of the weak-lensing community will be important for future weak-lensing surveys. We explore the impact that undersampling (\S \ref{sec:undersampling}), the brighter-fatter effect (\S \ref{sec:BFE}), focus variations of the telescope (\S,\ref{sec:focus}), and the slope of the galaxy SED (\S,\ref{sec:sed}) can have on the PSF and weak-lensing measurements. The conclusions from this work are critical for any future NIR weak-lensing (NIRWL) analysis that will detect mass overdensities in the CANDELS fields (Finner et al. in preparation). We develop PSF models for the F160W filter and test their validity in Sections \ref{sec:PSF} and \ref{sec:validation}, respectively. These PSF models will be used in our upcoming weak-lensing analysis for NIRWL.



In this work, magnitudes are reported in the AB magnitude system \citep{1974oke}. 

\section{Observations} \label{sec:obs}
CANDELS is one of the widest available HST/WFC3-IR surveys with a coverage of 0.23 square degrees \citep{2011koekemoer}. The excellent
resolution and sensitivity of HST/WFC3 yields a large number density of resolved galaxies ($\mytilde60$ galaxies arcmin$^{-2}$) to a $5\sigma$ magnitude limit of \mytilde27. This makes CANDELS the optimal data set to investigate NIR weak lensing for future space missions.

\subsection{CANDELS Fields} \label{sec:candels}
The CANDELS observations cover five fields: Great Observatories Origins Deep North and South (GOODS-N and GOODS-S), the UKIDSS Ultra-Deep Survey (UDS), the Extended Groth Strip (EGS), and the Cosmic Evolution Survey (COSMOS). Observations in these fields were achieved in multiple optical and NIR bands with HST. \cite{2011koekemoer} give a detailed account of the observations and the original processing of the CANDELS data. However, our goal is to achieve a high-fidelity weak-lensing analysis. Therefore, we selected the F160W-filter imaging because it is the deepest of the CANDELS IR imaging and has the best sampling rate (PSF FWHM to pixel scale ratio) with a PSF FWHM \mytilde0\farcs18 and pixel scale \mytilde 0\farcs13. The remainder of this study is focused on F160W imaging unless explicitly mentioned. 

\subsubsection{Astrometry Correction} \label{sec:astrometry}
The recent release of the Gaia DR3 catalog \citep{2022gaia} provides an opportunity to improve the astrometry of the CANDELS mosaics. When comparing the positions of Gaia stars in the magnitude range $18 < $ Gaia r-band $ < 21.5$ to their counterparts in the CANDELS Treasury mosaics v1.0\footnote{https://archive.stsci.edu/hlsp/candels} \citep{2011grogin, 2011koekemoer}, we discovered significant astrometric offsets. 
The average offset per field of the Treasury mosaic star positions from proper motion corrected Gaia positions vary from 0\farcs07 to 0\farcs26 (Table \ref{tab:alignment}). In the case of the UDS, the offset systematically varies from about 0\farcs1 in the center of the field to 0\farcs25 near the edge. In contrast, the GOODS-S field has a nearly constant 0\farcs26 offset (or shift) in relation to Gaia. An in-depth look at the variations in astrometric offset are presented in Appendix \ref{app:candels_redo}. 

\begin{table}[b]
\caption{Alignment of CANDELS mosaics} \label{tab:alignment}
\def\arraystretch{1.}
\centering
\begin{tabular}{ c  c  c  }
\hline
\hline
Field & Treasury Alignment [\arcsec] & Our Alignment [\arcsec]           \\
\hline
UDS & 0.13 $\pm$ 0.05 & 0.01 $\pm$ 0.03             \\
GOODS-S & 0.26 $\pm$ 0.05  & 0.02 $\pm$ 0.02            \\
GOODS-N & 0.05 $\pm$ 0.03 & 0.02 $\pm$ 0.02 \\
EGS & 0.07 $\pm$ 0.05 & 0.03 $\pm$ 0.02            \\
COSMOS & 0.11 $\pm$ 0.04 & 0.02 $\pm$ 0.03 \\
\hline
\end{tabular}

\end{table}

New observations that overlap with the CANDELS fields have been taken since the release of the data in \cite{2011koekemoer}. To reach the maximum coverage and depth, we have included the new observations that overlap significantly with the original footprint of the CANDELS fields. The HST/WFC3 observations that are used in this study are listed in Table \ref{tab:sci_obs}. The inclusion of newer observations increases the total footprint of the CANDELS fields by 17\%.

\begin{table}[]
\caption{Observations Included in This Study} \label{tab:sci_obs}
\def\arraystretch{1.}
\centering
\begin{tabular}{ c  c  c  c }
\hline
\hline
Field & PI & PROPOSIDs & Num. Frames           \\
\hline
GOODS-S & Faber & 12060/1/2 & 356             \\
GOODS-S & Illingworth & 11563 & 144            \\
GOODS-S & Riess & 12099/12461 & 28 \\
GOODS-S & Ellis & 12498 & 52            \\
GOOOS-S & Swinbank & 12866 & 4            \\
GOOOS-S & O'Connell & 11359 & 60            \\
\hline
GOODS-N & Faber & 12443/4/5 & 387             \\
GOODS-N & Riess & 12461/13063 & 52             \\
GOODS-N & Oesch & 15977 & 16             \\
GOODS-N & Ono & 12960 & 8             \\
\hline
UDS & Faber & 12064 & 176             \\
UDS & Ouchi & 12265 & 56             \\
UDS & Riess & 12099 & 31             \\

\hline
EGS & Faber & 12063 & 176             \\
EGS & Riess & 12099/13063 & 28             \\
EGS & Kocevski & 13868 & 4             \\
EGS & De Barros & 15103 & 18             \\

\hline
COSMOS & Franx & 12167 & 8             \\
COSMOS & Faber & 12440 & 177             \\
COSMOS & Riess & 12461 & 23             \\
COSMOS & Muzzin & 12990 & 4             \\
\hline

\end{tabular}

\end{table}

Our alignment of the frames was achieved in a multi-step process. First, frames taken within a single orbit (2-4 frames typically) were relative aligned by cross-matching stars and performing an iterative grid search to optimize the alignment. The grid search converged when a relative alignment of $0\farcs01$ was found. The image headers were updated to include the optimal shift and the component frames of the orbit were coadded with Astrodrizzle \citep{2010fruchter, 2012gonzaga} to an \textit{orbit patch} (consisting of frames taken within an orbit). Next, the \textit{orbit patches} that significantly overlap (header CRVAL within 60 arcseconds) were passed through the same grid search pipeline, an optimal alignment was found, and the headers of the component frames were updated, in unison, to the new alignment. These frames were drizzled into a \textit{group patch} (consisting of multiple \textit{orbit patches}). 

The last step was an absolute alignment. To achieve the purest catalog of stars for absolute alignment, we selected a sample of bright, unsaturated, and isolated stars from the CANDELS Treasury mosaics, as described in Appendix \ref{app:star_selection}. We used the Treasury stars to select stars in the \textit{group patches} and created catalogs of robust star positions that correspond to each \textit{group patch}. For each \textit{group patch}, we cross-matched the star catalog with the proper-motion-corrected Gaia catalog, which left about 0-5 unsaturated stars per patch. To alleviate the lack of stars, a hybrid reference catalog was created by combining Gaia stars with a supplementary catalog. When available, we chose the HSC SSP DR3 catalog \citep{2022aihara} as a supplement. However, wherever the HSC had no coverage, we relied on the Pan-STARRS DR2 catalog \citep{2020flewelling}. Both of these supplementary catalogs are aligned to Gaia. The hybrid reference catalog provides 3-15 unsaturated stars per \textit{group patch}. Using the hybrid reference catalog, a final iterative search for the best alignment of the stars in the \textit{group patches} was performed. The optimal alignment for each \textit{group patch} was applied to all FLT files that composed the patch to update their astrometry to the final alignment.

\subsubsection{Image Coaddition} \label{sec:coadd}
There are systematic effects in the imaging that should be corrected before image coaddition. Before coadding the images, we investigate the systematics and correct for them where possible. The CANDELS frames contain spurious signals from stray light, persistence, and satellites as well as comet tails from drift during integration. Each of these signals can affect a weak-lensing measurements and must be corrected. Observations that contain comet tails were discarded because their signal would bias weak-lensing shapes. Satellite trails were detected by visual inspection of the images and then masked. 

Persistence is the residual signal that remains in the detector after a reset and is common for HgCdTe detectors. For WFC3-IR, \cite{2012long} showed that persistence is non-linear and arises in previously saturated pixels. This is an issue for weak lensing because the persistent signal could lead to falsely detected sources or could affect the measurements of galaxy shapes. The HST website\footnote{https://archive.stsci.edu/prepds/persist/search.php} provides diagnostic images that contain the expected amount of persistence for each pixel of imaging. We utilized these images and masked any pixel that contains a persistence level above 0.01 electrons per second. In general, the levels of persistence in the CANDELS images are low with less than one percent of pixels lost per frame.

Stray light introduces a significant gradient in the imaging background, especially in the UDS field. To remove the gradient, we created a background model with SExtractor \citep{1996bertin}. Our primary goal was to subtract the stray light without impacting the galaxies. Therefore, we chose a SExtractor background size of 128 pixels and DETECT\_THRESH\footnote{https://sextractor.readthedocs.io/en/latest/Introduction.html} of 0.5. The background image was subtracted from the FLT image. For bright and extended sources, over-subtraction leads to large flux cavities \citep[for example see][]{2019aihara}. To avoid over-subtraction, we inspected the wings of the radial profile of bright sources in the background subtracted image. If the light profile of the wings was within the root mean square (rms) level of the background, we accepted the background subtraction. In some cases, we were unable to properly model the stray light without causing over-subtraction. In those cases, we discarded the frames.

After re-aligning the frames and removing the spurious signals mentioned above, we stacked the component frames into a mean-coadded mosaic with Astrodrizzle. We utilized a Gaussian kernel with a PIXFRAC\footnote{https://drizzlepac.readthedocs.io/en/latest/astrodrizzle.html} of 0.7 and a PIXSCALE of 0\farcs05. These coadded mosaics are the final data product that will be used for our future weak-lensing measurements. The newly drizzled products vastly improve the absolute astrometry of the five CANDELS mosaics. Comparing the proper-motion-corrected positions of stars in the Gaia DR3 catalog with their counterparts in the new mosaics shows that the average median offset is $0\farcs02$ or approximately 1/6th of the native pixel scale (Table \ref{tab:alignment}). In the case of the UDS, GOODS-S, and COSMOS, the new absolute alignment is a vast improvement over the Treasury mosaics. The mosaics that we have generated are available to download\footnote{\href{https://drive.google.com/drive/folders/1k9WEV3tBOuRKBlcaTJ0-wTZnUCisS__r?usp=share_link}{Link to download NIRWL mosaics.}}.

For our PSF modeling, an additional processing step is required. Each CANDELS frame was single drizzled to the native detector orientation using Astrodrizzle parameters that are consistent with the mosaics. Drizzling to the native detector orientation simplifies the pixel-to-pixel comparison of the frames to archival globular cluster imaging (Section \ref{sec:ngc104}), which is the reference dataset that we use to model the PSF (Section \ref{sec:PCA}). 

\begin{figure}
    \centering
    \includegraphics[width=0.5\textwidth]{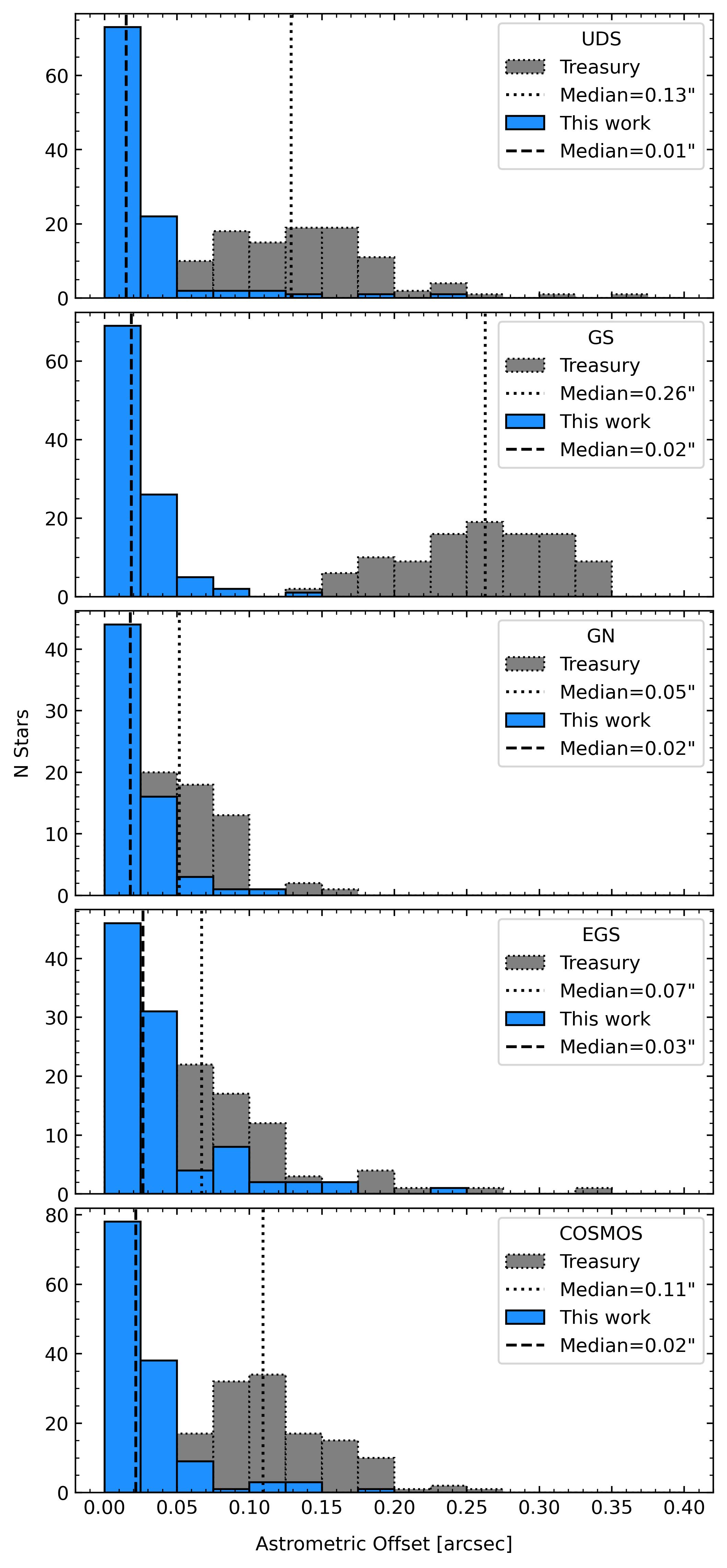}
    \caption{Astrometric offset of the CANDELS Treasury mosaics v1.0 (gray) and the mosaics produced in this work (blue) from the positions of stars in the proper-motion-corrected Gaia catalog DR3. Vertical dashed (this work) and dotted (Treasury) lines mark the median astrometric offset. On average, our updated mosaics have astrometric offset from the Gaia catalog of 0\farcs02, a factor of several improvement compared to the Treasury mosaics.
    }
    \label{fig:astromety_hist}
\end{figure}

\subsection{NGC 104} \label{sec:ngc104}
The globular cluster NGC 104 is an excellent field to study the telescope PSF because of its dense star field. 
We downloaded the NGC 104 WFC3-IR frames from the MAST archive \footnote{https://mast.stsci.edu/portal/Mashup/Clients/Mast/Portal.html} (summarized in Table \ref{tab:ngc104_obs}). These NGC 104 observations were selected because they target the outskirts of the globular cluster and thus provide a 
spatial density of stars that permit modeling the PSF to a radius of \mytilde0\farcs75 without significant contamination from neighbors.

\begin{table}[ht]
\caption{NGC 104 Observations} \label{tab:ngc104_obs}
\def\arraystretch{1.}
\centering
\begin{tabular}{ c  c  c }
\hline
\hline
PI & PROPOSIDs & Num. Frames           \\
\hline
Richer & 11677 & 224             \\
\multirow{2}{*}{ Hilbert} & 11453, 11931, 12352,  & \multirow{2}{*}{ 118}             \\
& 12696, 13079, 13563 & \\
Dressel & 11445 & 15             \\
Riess & 14868 & 3             \\
\hline

\end{tabular}

\end{table}

As done with the CANDELS frames, we single drizzled the NGC 104 frames with consistent Astrodrizzle parameters and to the native detector orientation. These single drizzled NGC 104 frames are used extensively in the following systematics study (Section \ref{sec:systematics}) and in the generation of PSF models (Section \ref{sec:PCA}).

\section{PSF Systematics of HST/WFC3-IR}\label{sec:systematics}

Systematic effects from the light passing through the optics of the telescope and those arising from the sensing of the light by the detector can significantly impact a weak-lensing analysis. These effects manifest in the point-spread function (PSF) and the pixel-response function (PRF), respectively. For weak lensing with the HST Advanced Camera for Surveys (ACS), the former effects have been well-studied in the optical regime \cite[see][]{2007jee, 2007rhodes}. However, a robust understanding of the PSF and PRF for weak lensing with the WFC3-IR detector is still needed. In this section, we investigate the most significant systematic effects that will impact a weak-lensing study with WFC3-IR.
 
 \begin{figure*}
    \centering
    \includegraphics[width=\textwidth]{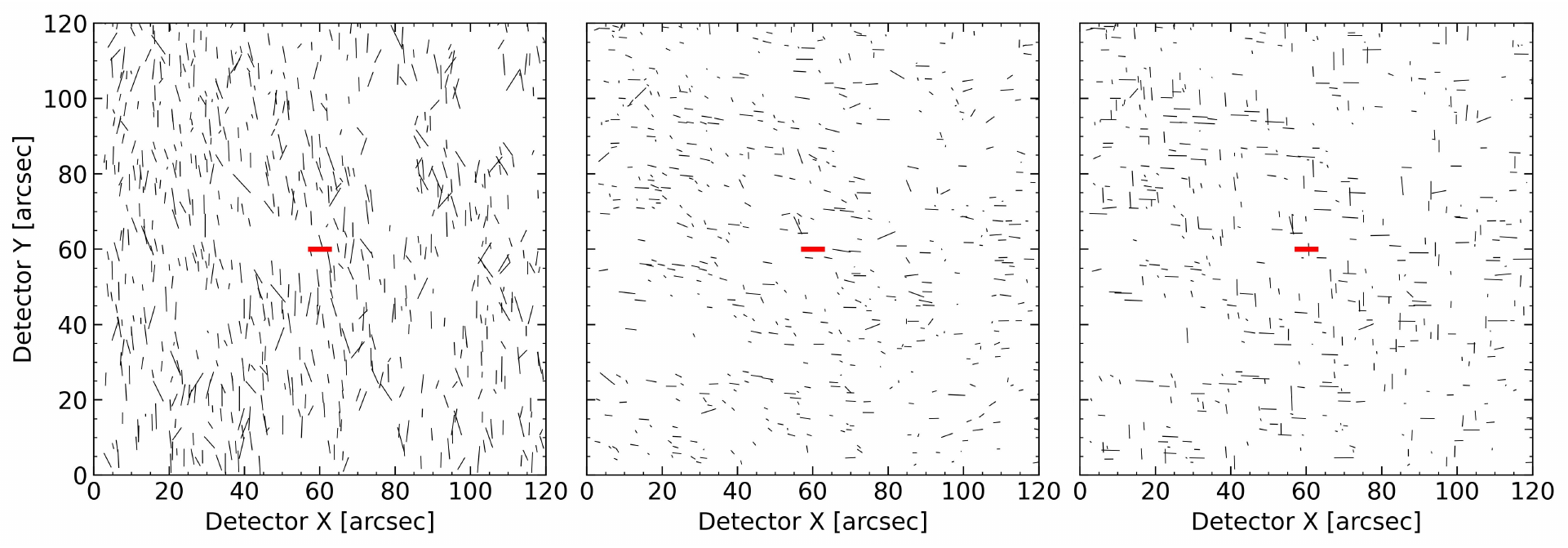}
    \caption{Magnitude and direction of the ellipticity of stars from a single NGC 104 frame observed with the F160W filter. The red line represents an ellipticity of $e_1=0.05$, $e_2=0$}. \textit{Left}: Stars from an FLT frame, which has not been corrected for geometric distortion. Geometric distortion causes the \textit{y}-axis elongation of the PSF. The plate scale is approximated as 0.13 arcsec/pixel for display purposes. \textit{Middle:} Stars from the same frame after the geometric distortion is corrected by Astrodrizzle. The drizzled PSF elongates in the \textit{x} direction and the ellipticity is smaller than in the FLT frame. \textit{Right:} An improper weight function applied when measuring the shapes of stars in the drizzled frame emphasizes aliasing, which causes the ellipticity of stars to systematically align with the \textit{x} and \textit{y} detector axes.  
    
    \label{fig:flt_psf}
\end{figure*}

Throughout this work, we utilize the quadrupole moments to characterize the size and shape of objects. The first moment or centroid is defined as
\begin{equation} \label{eq:centroid}
    \bar{x}_{i,j}=\frac{\int d^2x\ I(\textbf{x}) x_{i,j} }{\int d^2x\ I(\textbf{x})}.
\end{equation}
The quadrupole moments are measured with respect to the centroid. However, in most cases in this work (unless explicitly stated), we utilize the SExtractor measured centroid (XWIN\_IMAGE, YWIN\_IMAGE) rather than Equation \ref{eq:centroid} because it is less affected by neighboring objects. The quadrupole moments are defined as

\begin{equation} \label{eq:quadrupole}
    Q_{ij} = \frac{\int d^2x\ I(\textbf{x})W(\textbf{x})(x_i-\bar{x}_i)(x_j-\bar{x}_j)}{\int d^2x\ I(\textbf{x})W(\textbf{x})},
\end{equation}
where $I(\textbf{x})$ is a 2D image, $W(\textbf{x})$ is a weight function, and $i,j$ are 1 or 2 to signify the two Cartesian spatial dimensions. The choice of a weight function is important to the stability of the measurement \citep{2021kannawadi}. We select a circular 
Gaussian weight function and set its standard deviation to $\sigma_w=0\farcs2$. This choice sufficiently reduces noise in the quadrupole measurement while remaining large enough to capture the shape of the WFC3-IR F160W PSF, which has a full width at half maximum (FWHM) of \mytilde$0\farcs18$.

From the quadrupole moments, the complex ellipticity of an object is defined as:
\begin{equation}
    e_1+ie_2 = \frac{Q_{11}-Q_{22}+2iQ_{12}}{Q_{11}+Q_{22}+2\sqrt{Q_{11}Q_{22}-Q_{12}^2}}.
\end{equation}
This ellipticity is equivalent to the ellipticity (a-b)/(a+b), where a and b are the semi-major and semi-minor axes of the ellipse. The size of an object is defined as
\begin{equation}
    R = \sqrt{Q_{11}+Q_{22}}.
\label{eq:quad_size}
\end{equation}

Figure \ref{fig:flt_psf} shows the quadrupole measurement of stars in a single NGC 104 frame. The whiskers (black lines) represent the magnitude and direction of the ellipticity of the stars or equivalently the effective PSF (the PSF convolved with the PRF). In the left panel, are the measurements of the stars in an FLT frame. An FLT frame is a flat-fielded product that has not been corrected for geometric distortion. The geometric distortion is caused by the tilt of the focal plane with respect to the incoming beam, which projects a square into a rectangle and imparts a \textit{y}-direction elongation to the PSF \citep{2017jee}. The FLT frames will not be used for modeling the PSF but they do represent the imaging at the focal plane and are a useful tool for studying systematics. The middle panel shows the ellipticity of the PSF in a drizzled frame after correcting for geometric distortion with Astrodrizzle. Here, the PSF is elongated in the \textit{x} direction with a smaller magnitude than the distorted frame. In comparison to the ACS PSF, the WFC3-IR PSF does not show the smooth, but significant, spatial variation in size and ellipticity \citep[for ACS see][]{2007jee, 2007rhodes}. Another advantage over the ACS is that the WFC-IR detector is a single chip and does not suffer from a discontinuity at the chip boundary that is found in some multi-chip detectors. In these aspects, the WFC3-IR PSF is simpler than its ACS counterpart. 

To understand how systematic effects impact a weak-lensing analysis, one must put them into the context of weak lensing. Two of the image distortions of weak lensing are the convergence and shear. The convergence acts isotropically and magnifies galaxies while preserving surface brightness. The shear is anisotropic and modifies the ellipticity of galaxies. In weak-lensing analysis, the shapes of galaxies are measured and then averaged to recover the shear. However, all shape measurement techniques are subject to biases such as noise bias, model bias, truncation bias, etc \citep[for more on biases see][]{2018mandelbaum}. To arrive at an accurate and precise measurement, it is desirable to minimize biases and, when required, to properly account for them. The contribution of the biases to the measurement of the shear can be separated into a multiplicative bias (\textit{m}) and an additive bias (\textit{c}) as follows:

\begin{equation} \label{eq:bias}
    g_m-g_t = mg_t + c,
\end{equation}
where $g_m$ is the measured shear and $g_t$ is the true shear \citep{2006huterer}. In this work, we refer to the bias on each galaxy measurement as a galaxy shape bias and the bias on the averaged ellipticity of an ensemble of galaxy images (measured shear) as a shear bias.

\subsection{Undersampling} \label{sec:undersampling}
The Nyquist-Shannon sampling theory \citep{Shannon1949} states that a signal must be sampled at a rate of at least twice its frequency to be reconstructed. In pixelated detectors, the condition for a critically sampled signal is approximately 2 pixels per FWHM of a Gaussian PSF \citep{2005puetter, 2017robertson}. At 1.6$\mu$m, the WFC3-IR PSF has an FWHM of approximately 0\farcs18, which is sampled with pixels of size $\mytilde0\farcs128$. Thus, the PSF is undersampled. Following \cite{2021kannawadi}, we define the sampling factor, $S$, as the PSF standard deviation to plate scale ratio. This sets the critical criteria to $S = 1$. Therefore, $S > 1$ is well sampled, $S<1$ is undersampled, and $S < 0.5$ is severely undersampled.

\begin{figure*}
    \centering
    \includegraphics[width=\textwidth]{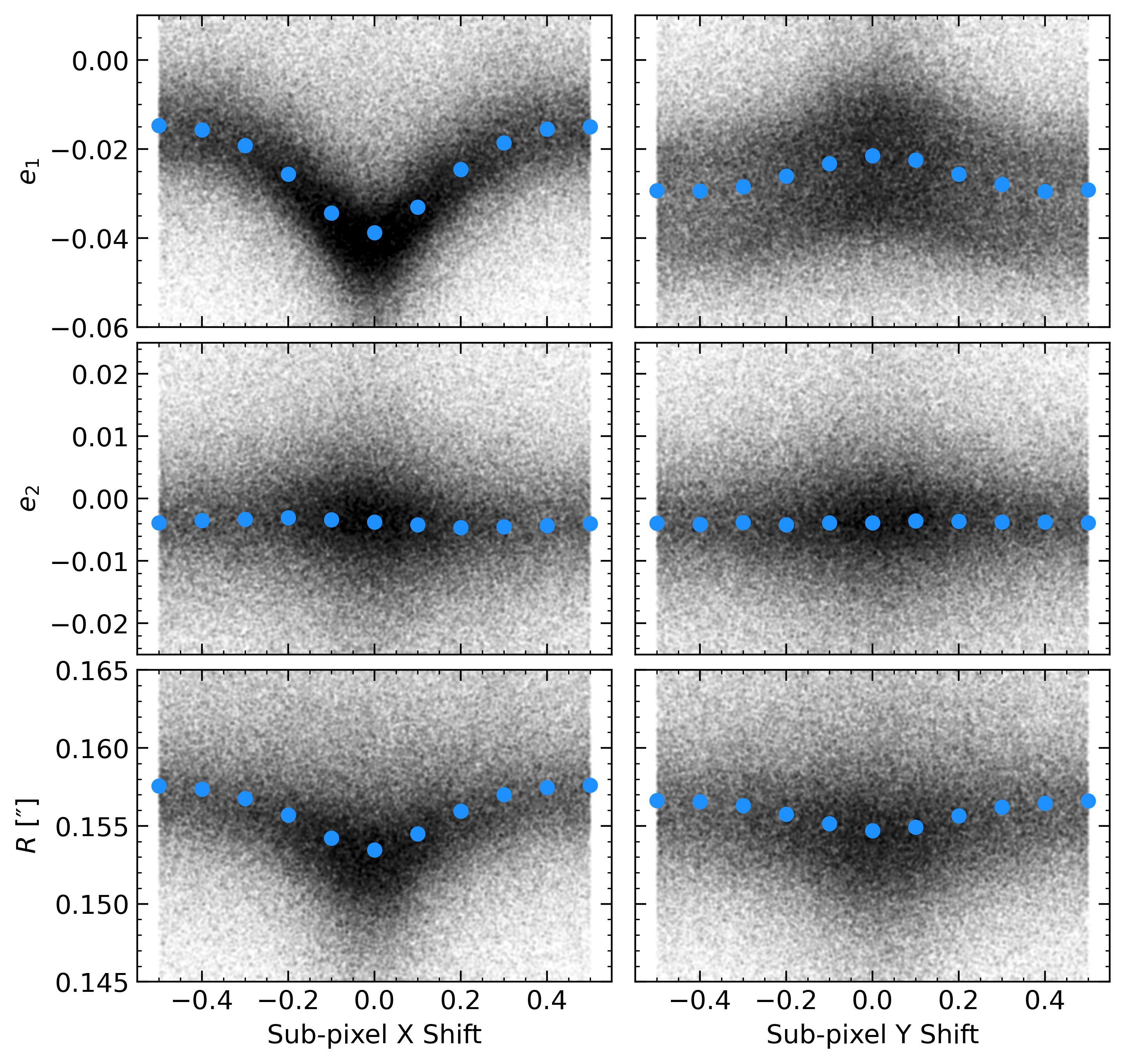}
    \caption{Observed sub-pixel centroid dependence of $e_1$ (row 1), $e_2$ (row 2), and $R$ (row 3) of stars in FLT frames of NGC 104. Black circles represent the quadrupole measurements of individual stars. Blue circles are median values with bin width of 0.1 pixels. The standard error ($\sigma/\sqrt{n}$) for each of the blue circles is $<0.01\%$. \textit{Row 1:} The image of stars (PSF) centered in a pixel have ellipticity in the \textit{y} direction. Sub-pixel \textit{x} shifts cause a systematic elongation of the PSF in the \textit{x} direction, which leads to a rounder PSF while sub-pixel \textit{y} shifts cause an elongation of the PSF in the \textit{y} direction. \textit{Row 2:} Sub-pixel shifts do not cause a discernible variation in the $e_2$ component. \textit{Row 3:} The size of the PSF increases for both \textit{x} and \textit{y} shifts away from the center of the pixel. Thus, there is a sub-pixel location dependency for the ellipticity and size of the PSF.}
    \label{fig:undersampling_flt}
\end{figure*}

One of the byproducts of undersampling is aliasing. Aliasing is the misinterpretation of a high-frequency signal as a lower-frequency signal. The right panel of Figure \ref{fig:flt_psf} shows a manifestation of aliasing in the direction and magnitude of the quadrupole moments of stars in the same frame that is presented in the middle panel. In this case, rather than weighting Equation \ref{eq:quadrupole} using a Gaussian of $\sigma_w = 0\farcs2$ when measuring the quadrupole moments, we used $\sigma_w=0\farcs12$. The smaller weight emphasizes the PSF undersampling, which leads to alignment of the PSF ellipticity along the axes defined by the pixel orientation.

To further understand the effect of undersampling on the shapes of objects, we need to investigate what occurs at the pixel level. The quantization of the light from stars and galaxies into square pixels modifies the shape of the image. In a well-sampled image, the change in shape caused by pixelization is a subdominant effect and the image well approximates the true shape. However, in the case of an undersampled image, the change in shape by pixelization can be severe and becomes dependent on the sub-pixel location of the image centroid.    

The undersampling effect and its sub-pixel dependence can be detected in the F160W images of NGC 104. The rows of Figure \ref{fig:undersampling_flt} present the quadrupole measured shapes of stars (black circles) as a function of sub-pixel location of the centroid. These stars were selected from the NGC 104 images prior to geometric correction (FLT images). We choose to present the measurements in the FLT images because they contain the original position of the object when undersampling occurred. The blue circles are the median values in 0.1 pixel bins with standard errors ($\sigma/\sqrt{N_{stars}}$). The left and right panels focus on the effect of sub-pixel position of the star centroids along the \textit{x} direction and \textit{y} direction, respectively.  

The intrinsic elongation of the PSF in the FLT frame, as shown in Figure \ref{fig:flt_psf}, is in the \textit{y} direction. This agrees with the top row of Figure \ref{fig:undersampling_flt}, where the average $e_1$ component is negative. The left panel shows that a star (the PSF) with a centroid that is displaced in the \textit{x} direction elongates in the \textit{x} direction. In this case, the observed PSF becomes rounder because $e_1$ changes from approximately -0.04 to -0.015. In the right panel, we find that displacements in the \textit{y} direction cause elongation in the \textit{y} direction which decreases the $e_1$ component. However, the magnitude of the change is lower than for \textit{x} displacements with \mytilde 0.01 decrease. As we show in the simulation in the following subsection, the asymmetry of the effect is caused by the intrinsic shape of the PSF.

In the second row, there is minimal variation in the $e_2$ component. Thus, undersampling does not significantly modify the shape of the PSF in the diagonal direction of a pixel. 

The third row shows the dependence of the PSF size with sub-pixel position of the centroid. We find that both \textit{x} and \textit{y} displacements increase the size of the PSF with x displacements having more impact on the size of the PSF than y displacements. The largest increase in median size is approximately $3\%$. We reiterate that these measurements have all been performed in the geometrically distorted FLT frames. The corrected (drizzled) frames have an intrinsic ellipticity that runs in the \textit{x} direction. We present our investigation of undersampling in the drizzled frames in Appendix \ref{app:undersampling}. The investigation finds similar patterns and magnitudes as found for the FLT frames.

\begin{figure*}
    \centering
    \includegraphics[width=\textwidth]{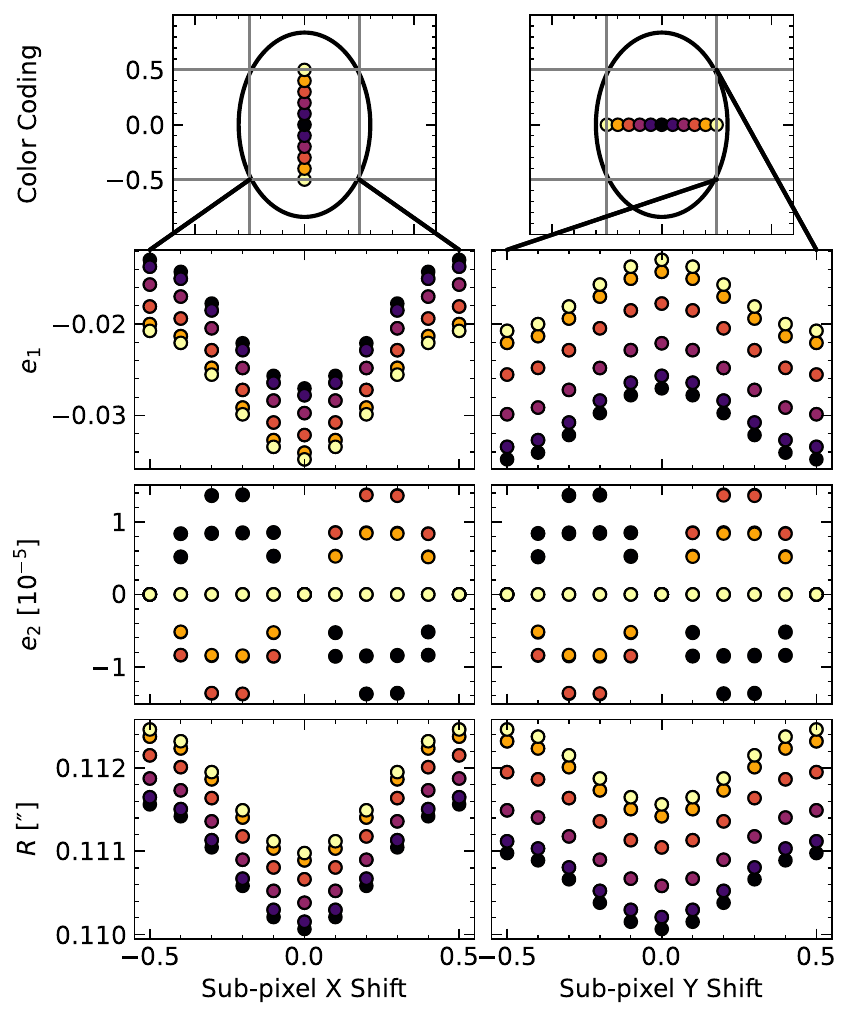}
    \caption{Simulated undersampling of an elliptical Gaussian distribution with $e_1=-0.03, e_2=0$ and a sampling rate \textit{S}=0.6. \textit{Row 1:} The color scheme showing the sub-pixel dithers along the off axis, where black represents no sub-pixel shift and yellow is the maximum sub-pixel shift of 0.5 pixels. \textit{Row 2:} Shifts in the \textit{x} direction elongate the PSF in the \textit{x} direction, which circularizes the PSF. Shifts in the \textit{y} direction elongate the PSF in the \textit{y} direction making it more elliptical. The simulation highlights that the scatter in Figure \ref{fig:undersampling_flt} is partly caused by shifts along the opposite axis. \textit{Row 3:} As in the observed PSF, the $e_2$ component is not significantly affected by shifts. \textit{Row 4:} The size of the PSF with both \textit{x} or \textit{y} shifts enlarges the PSF.}
    \label{fig:undersampling_sim}
\end{figure*}

\subsubsection{Simulating Undersampling}
To gain further insight into the effect of undersampling on the ellipticity and size of the PSF, we simulated a 2D Gaussian distribution (FWHM=0\farcs18) and sampled it at a rate of $S=0.6$ (approximately the value for HST/WFC3-IR F160W). To mimic the shape of the FLT PSF, the Gaussian was given an intrinsic ellipticity in the \textit{y} direction by setting $e_1=-0.03$ and $e_2=0.0$. The simulations stepped through sub-pixel dithers on the object centroid from -0.5 to +0.5 pixels in the \textit{x} and \textit{y} directions. 

Figure \ref{fig:undersampling_sim} shows the dependence of measured PSF parameters on sub-pixel centroid position for the shape of an undersampled Gaussian distribution. As in Figure \ref{fig:undersampling_flt}, the left and right panels present the \textit{x} and \textit{y} displacements of the centroid, respectively. Displacements along the axis that is not the focus of the panel are tracked by color (ie. y displacements in the left panels and x displacements in the right panels). This tracking scheme is graphically shown in the first row. 

In all rows, the simulations reflect what is observed from WFC3-IR F160W observations of NGC 104 with a similar range of variation. Therefore, we will not repeat these points. However, the simulations are able to separate the \textit{x} and \textit{y} displacements. They show that some of the scatter that was observed in Figure \ref{fig:undersampling_flt} is caused by displacements along the axis that is not the focus of the panel. 
Furthermore, the amount of scatter is dependent on the intrinsic ellipticity of the object. Since the simulated (and observed) PSF has an intrinsic ellipticity in the \textit{y} direction, the effect of x displacements cause a larger variation in the PSF shape. This is likely caused by the lower sampling rate along the minor axis of the ellipse. 
We verified this by simulating a circular Gaussian distribution and found that the effect of \textit{x} and \textit{y} displacements are symmetric. 
Ideally, if we knew the exact centroid of a source within the pixel, this effect could be modeled and calibrated out. But, due to detector effects such as quantum efficiency variation within the pixel, the true centroid of a source is never known and
it is better to mitigate the role of undersampling statistically.
\subsubsection{Impact of Filter on Undersampling}
In Appendix \ref{app:undersampling}, we present undersampling in four HST WFC3-IR filters: F105W, F125W, F140W, and F160W. As one would expect, the effect of undersampling on the PSF shape is a directly proportional to the sampling rate. 
Because the HST is diffraction limited, we find that the PSF in bluer filters (more undersampled) has more extreme changes in the ellipticity and size than in redder filters.

\begin{figure}[!ht]
    \centering
    \includegraphics[width=0.48\textwidth]{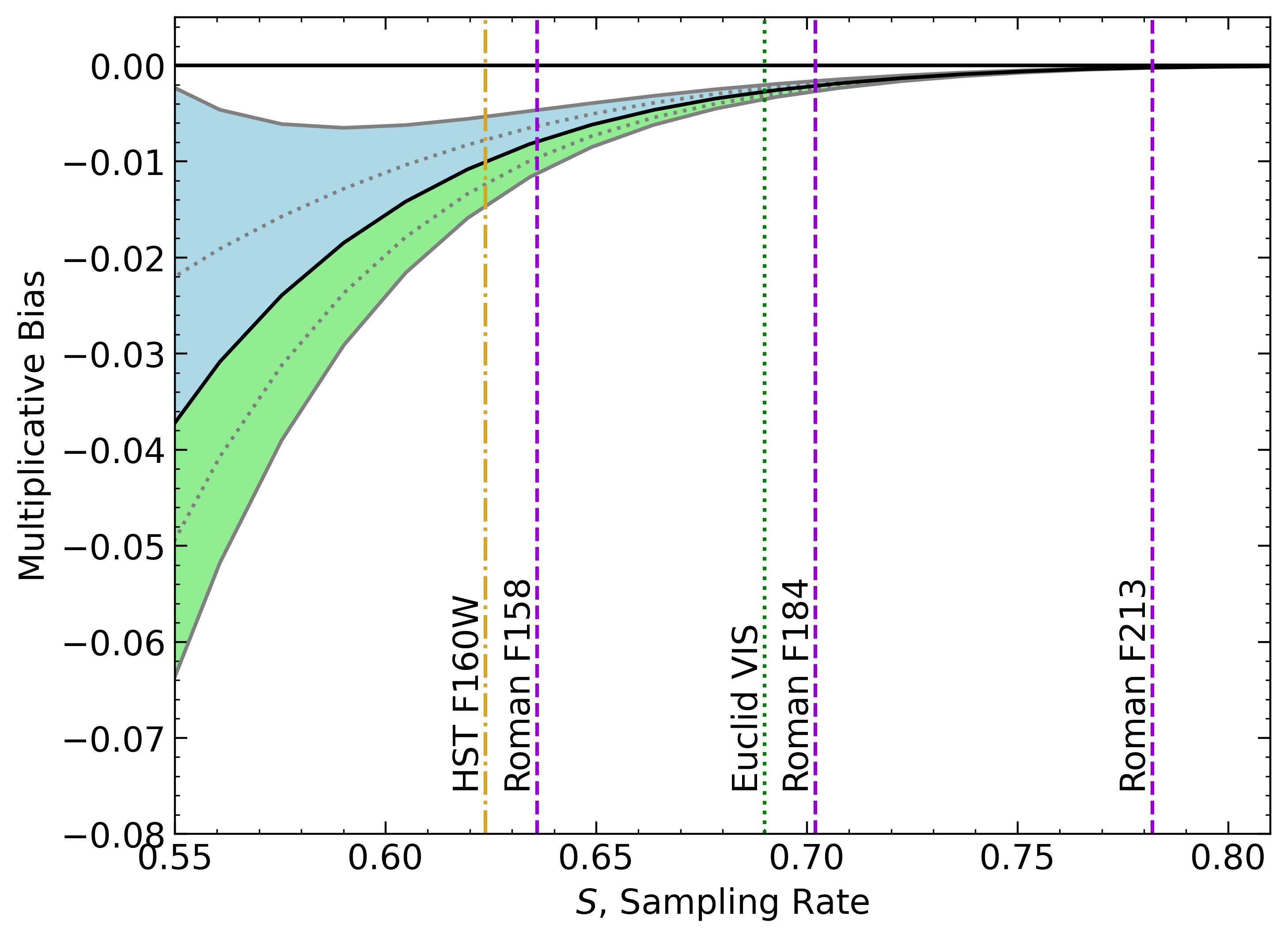}
    \caption{The effect of undersampling on galaxy shape measurement. The simulated galaxy has \textit{x}-direction elongation ($e_1 = 0.02$, $e_2 = 0.0$) and S\'{e}rsic parameters half-light radius=$0\farcs18$ and n=0.5. The galaxy has been convolved with a circular Gaussian PSF with FWHM=0\farcs2. The solid black curve traces the multiplicative galaxy shape bias caused by undersampling when the source is perfectly centered in a pixel. The shaded regions represent sub-pixel shifts of the galaxy centroid in the \textit{x} (blue) and \textit{y} (green) directions. The solid (dotted) gray lines mark sub-pixel shifts of 0.5 (0.25) pixels. Vertical lines indicate the sampling rate of notable filters with labels attached. The Roman F213 suffers the least from undersampling bias because of its larger PSF. The sub-pixel dependence of the galaxy shape is small for Roman F184 and Euclid VIS. For HST F160W and Roman F158, the dependence leads to a multiplicative bias variation of \mytilde0.012. The Euclid NISP filter has a sampling rate of \mytilde0.43 and will have severe bias from undersampling.}
    \label{fig:undersampling_bias}
\end{figure}

\subsubsection{Undersampling Effect on Shape Measurement} \label{sec:undersamp_sim}
It is important to understand the sampling effect on the PSF, but perhaps what is more important is to understand how sampling biases galaxy shape measurements. Due to the size evolution of galaxies with redshift, a significant fraction of the galaxies that are useful for weak lensing have characteristic sizes less than 0.3\arcsec. These galaxies tend to be high redshift and carry the majority of the lensing signal but they are prone to undersampling. To investigate the effect of undersampling on shape measurements, we performed a simulation with our shape-fitting pipeline. 

For simplicity, we assumed a S\'{e}rsic galaxy and a Gaussian PSF. The goal is to track multiplicative galaxy shape bias (Equation \ref{eq:bias}) of the galaxy shape measurement with sub-pixel position of the galaxy centroid. Using Galsim \citep{2015rowe}, we created a postage stamp image of a galaxy (half-light radius=0.18\arcsec and S\'{e}rsic index n=0.5) with an ellipticity of $e_1=0.02$ and $e_2=0.0$ and convolved it with a circular Gaussian PSF (FWHM=0.2\arcsec). We sampled the simulated image at sub-pixel displacements with a plate scale ranging from 0\farcs11 to 0\farcs22 per pixel. We then forward modeled the same PSF and fit a S\'{e}rsic distribution to the postage stamp image with the position fixed to the true sub-pixel centroid location. 

Figure \ref{fig:undersampling_bias} shows the multiplicative galaxy shape bias as a function of the sampling rate. The black line traces the effect of undersampling on the shape measurement for a galaxy that is centered in a pixel \cite[similar to what is presented in][]{2021kannawadi}.  We see that \textit{x} and \textit{y} displacements cause different bias than a galaxy that is centered on a pixel. A galaxy with centroid displaced in the \textit{x} direction (blue region) has less bias than the centered version because the undersampled galaxy is elongated in the intrinsic-shape direction. On the other hand, a galaxy with centroid displaced in the \textit{y} direction (green region) has more bias than the centered version because the elongation caused by the displacement is counteracting the intrinsic ellipticity of the galaxy.

The simulation presented in Figure \ref{fig:undersampling_bias} is only for a galaxy of a single size and ellipticity. However the simulation provides critical insight into how the sampling rate and the size of the PSF can affect galaxy shape measurement. To demonstrate the effect for other telescopes, we have included the sampling rate of the PSFs of HST, Euclid, and Roman filters on the figure. It is useful to make a relative comparison of the filters. We see that HST F160W has the lowest sampling rate of the presented filters. The Roman F158 filter will have comparable undersampling to HST F160W. Euclid VIS and Roman F184 are expected to be less affected by undersampling and Roman F213 will be the least affected. 

To estimate the effect that undersampling can have on a shear measurement, we simulate the population of galaxies from the UDS field using GalSim. We create a catalog of measured S\'{e}rsic parameters for the galaxies in the UDS field by fitting each with a psf-convolved S\'{e}rsic profile, where modeled PSFs from Section \ref{sec:PSF} are used. In GalSim, we generate 40,000 analytic galaxies (20,000 unique and 20,000 rotated $90^\circ$ counterparts for a ring test) that follow a bulge and disk model with an intrinsic ellipticity that is sampled from the UDS measurements. We randomly sample from the UDS distribution of fitted S\'{e}rsic parameters to create a disk component and add to it a bulge that has a fixed S\'{e}rsic index of four. The two components are combined with a bulge-to-total ratio randomly sampled from a Gaussian distribution with mean of 0.3 and standard deviation of 0.3 (keeping only positive values). 

\begin{figure}
    \centering
    \includegraphics[width=0.48\textwidth]{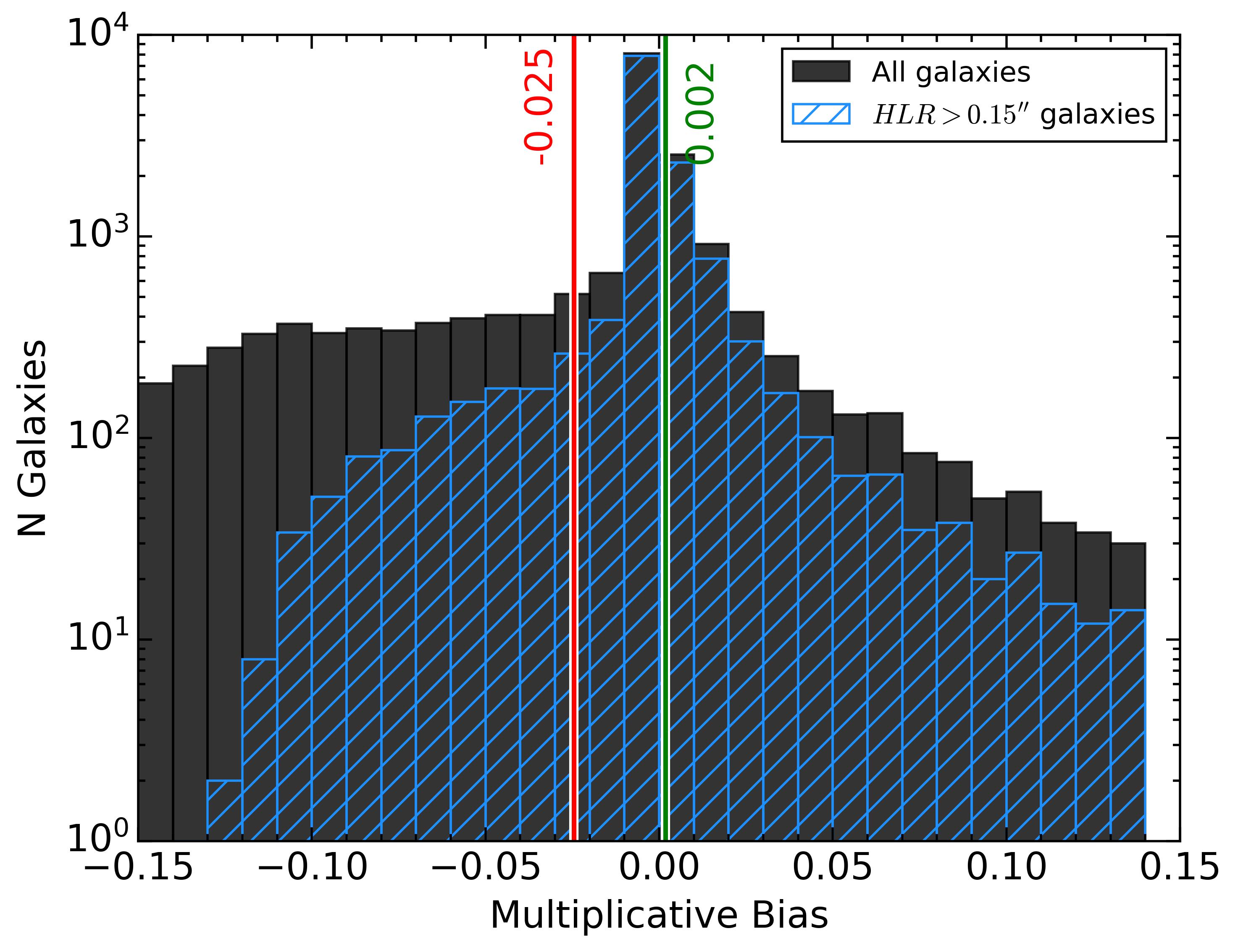}
    \caption{Histogram of galaxy shape bias for the population of galaxies from the UDS field. The vertical red line shows the multiplicative shear bias (-0.025) calculated from averaged galaxy ellipticities for applied shears in the range of -0.05 to 0.05 (see Section \ref{sec:undersamp_sim} for details of simulation). To demonstrate one method to mitigate the undersampling effect, we present a half-light radius ($HLR$) cut $> 0\farcs15$ (blue hatched histogram). After the $HLR$ cut (vertical green line), the multiplicative shear bias is reduced to 0.002.}
    \label{fig:uds_ebias}
\end{figure}

To simulate observed galaxies, we interpolate the average UDS PSF with GalSim \texttt{InterpolateImage} and convolve it to each galaxy. The average UDS PSF is chosen to remove any variation in PSF model from affecting the result. A postage stamp image of each galaxy is then created at a super-resolution of 0\farcs01 per pixel and downsampled to the native WFC3-IR scale of 0\farcs13 per pixel to imitate the effect of an undersampled detector. While generating the postage stamp image, we set the sub-pixel position of each galaxy to a random uniform number in the range -0.5 to 0.5 in the native pixel scale (adjusted for the super-sampling rate). An undersampled PSF postage stamp is created in the same manner but with the centroid fixed to the center of a pixel. We then perform forward-model S\'{e}rsic fitting on the undersampled image and compare the measured galaxy ellipticity to the input ellipticity to recover the galaxy shape bias. Figure \ref{fig:uds_ebias} presents a histogram of the galaxy shape biases.

We repeat this process for applied shears in the range of -0.05 to 0.05 at intervals of 0.01 and calculate the average ellipticity of the 40,000 galaxies for each applied shear. A linear fit to the average ellipticities as a function of the applied shear returns a multiplicative shear bias of $-0.025$ (red vertical line in Figure \ref{fig:uds_ebias}) and an additive bias $<10^{-4}$. The brute-force method for lowering this bias is to remove small objects from the catalog. We find that restricting the half-light radius to above $0\farcs15$ (slightly larger than one pixel) can lower the multiplicative shear bias to about 0.002 (green vertical line in Figure \ref{fig:uds_ebias}). However, the half-light radius cut drops the number density of galaxies from \mytilde81 to \mytilde 55 galaxies arcmin$^{-2}$. We perform the same fitting procedure on the super-resolution images and find an insignificant level of multiplicative shear bias.

We will explore the effect of undersampling further in our upcoming weak-lensing analysis (Finner et al. in preparation) by comparing the weak-lensing signal recovered from shapes measured in F160W with those measured in ACS F814W \citep{2007leauthaud}. We will also search for correlation of the galaxy shape with the measured sub-pixel position of the galaxy centroid.

\subsection{Brighter-fatter Effect} \label{sec:BFE}
The brighter-fatter effect (BFE) is becoming a well-studied systematic effect. It is caused by the drift in electric field from the build up of charge in detector pixels \citep{2014antilogus}. As more charge builds in a pixel, the electric field increases the chance that the next charge will form in adjacent pixels. This leads to a non-linear radial bleeding of charges and an artificial increase in the size of objects that is dependent on the accumulation of signal.

\begin{figure}
    \centering
    \includegraphics[width=0.48\textwidth]{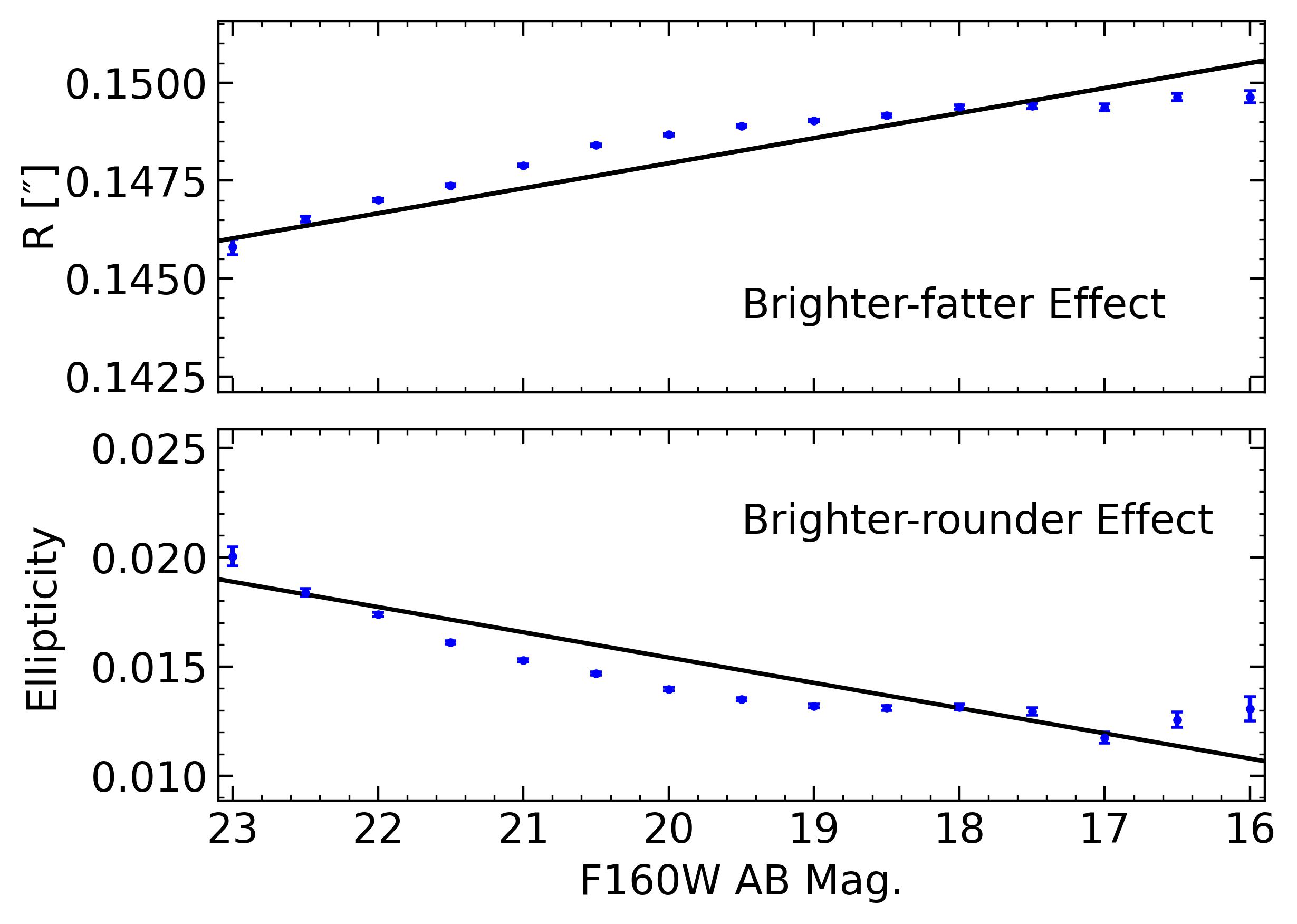}
    \caption{Quadrupole measured ellipticity (top panel) and size (bottom panel) of NGC 104 stars as a function of F160W AB magnitude. Blue circles are the median values in bins that are one magnitude wide with standard errors. The black line is a linear fit to the unbinned data. The data shows that brighter stars are observed to be rounder (top panel) and brighter (bottom panel). Across the magnitude range the ellipticity changes by \mytilde0.006 and the size changes by \mytilde2\%.}
    \label{fig:bfe}
\end{figure}

We tabulated all stars with $S/N > 5$ from the F160W frames of NGC 104. The top panel of Figure \ref{fig:bfe} shows the average size of stars, \textit{R} (defined in Equation \ref{eq:quad_size}), in 0.5 magnitude bins (blue circles). The blue error bars are the standard errors. The size-magnitude relation of stars is positively sloped with brighter ones being larger. A linear fit to the unbinned stars finds that the brightest, unsaturated stars are $\mytilde2\%$ larger than the faintest stars in our sample. \cite{2020finner} simulated the BFE by forward modeling an oversized PSF during Gaussian model fitting and found that a 5\% overestimation of the PSF size led to a multiplicative bias of 0.02.

The BFE is not the only effect that has a dependence on brightness. The bottom panel of Figure \ref{fig:bfe} shows the average ellipticity of stars in 0.5 magnitude bins. By a linear fit (black line) to the stars, we find a decrease in ellipticity by \mytilde0.006 from the faint to bright stars. We call this the brighter-rounder effect (BRE). We suspect that the BRE may stem from the isotropy of the BFE. If the charges that were expected to be detected in the central pixel have equal likelihood of being found in any of the adjacent pixels, the result should be a rounding of objects that is caused by the BFE. An alternative explanation could be that undersampling is playing a role. The BFE will effectively increase the sampling rate of brighter stars. The increased sampling rate would lead to less change in ellipticity for sub-pixel shifts (as presented in Section \ref{sec:undersampling}). Therefore, the brighter stars could have less undersampling induced ellipticity.

\begin{figure*}
    \centering
    \includegraphics[width=\textwidth]{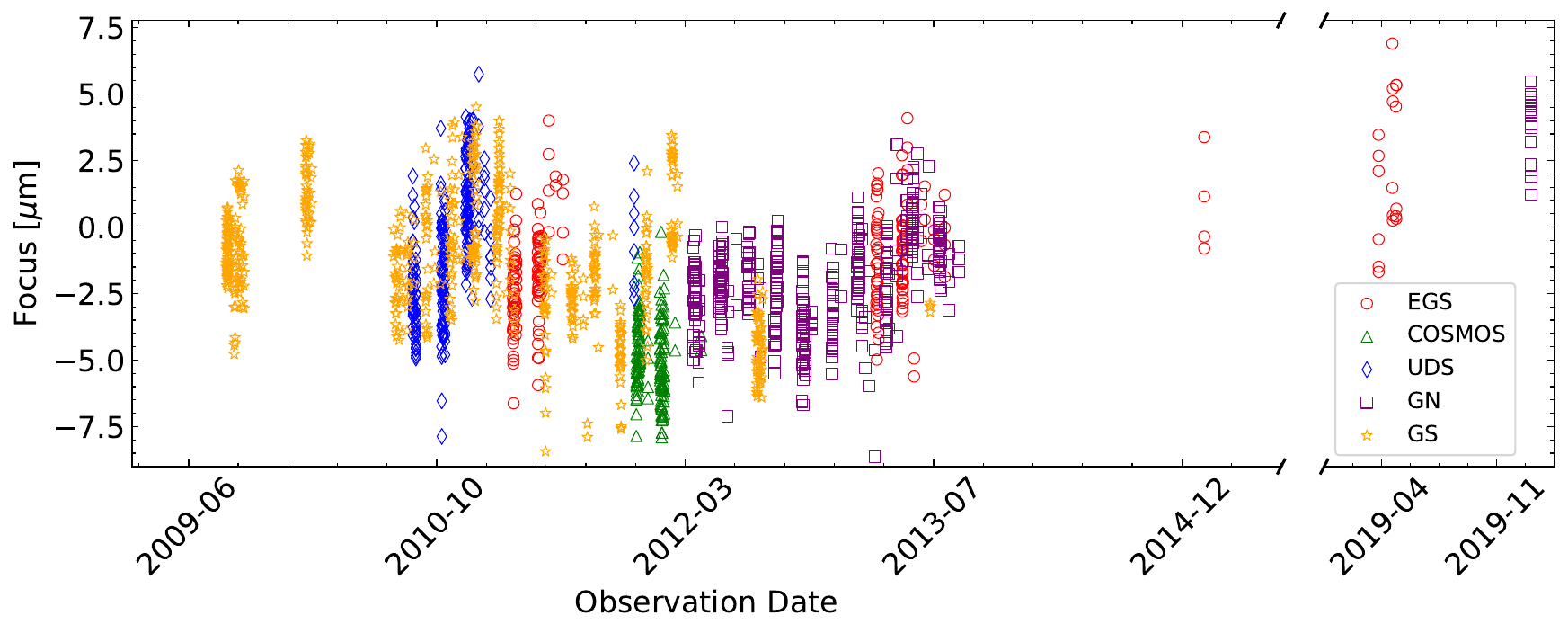}
    \caption{Temporal variation of the HST focus for observations in the CANDELS fields. The majority of observations are performed in the range of $-5 \mu m$ to $5 \mu m$. The temporal variation of the HST focus causes variations in the PSF that we account for in our PSF modeling (Section \ref{sec:PSF}).}
    \label{fig:model_focus}
\end{figure*}

\subsection{Focus Variations} \label{sec:focus}
A well-known systematic effect of the HST is the PSF variation caused by the changing focus. The HST is in a low-earth orbit that takes ninety minutes. 
During that period, the internal temperature of the telescope varies from the impingement of sunlight. The variation in temperature induces a variation in the focus of the optics. This is colloquially referred to as telescope ``breathing''. For the ACS, the breathing of the telescope causes significant variations in the PSF ellipticity and size \citep{2007jee, 2007rhodes}. The patterns shown in these works highlight that a well-modeled PSF must include the temporal variations. However, \cite{2016anderson} suggests that the longer wavelengths of the WFC3-IR are less susceptible to the focus variations. Nevertheless, we investigate the time dependence of the WFC3-IR PSF. 

\cite{2008dinino} have modeled the focus variation of the HST detectors as a function of average temperature sensor values. Their methods estimate the focus to a precision of $\sigma=0.5\mu m$. We retrieved the tables from the HST website\footnote{https://www.stsci.edu/hst/instrumentation/focus-and-pointing/focus/hst-focus-model} and cross-matched them with the average time of observation of each CANDELS frame. Figure \ref{fig:model_focus} displays the predicted focus for each CANDELS frame with colors differentiating the five fields. In general, the CANDELS observations were completed at telescope focus ranging from $-5\mu m$ to $+5 \mu m$ with some frames extending beyond this range. 

\begin{figure*}
    \centering
    \includegraphics[width=\textwidth]{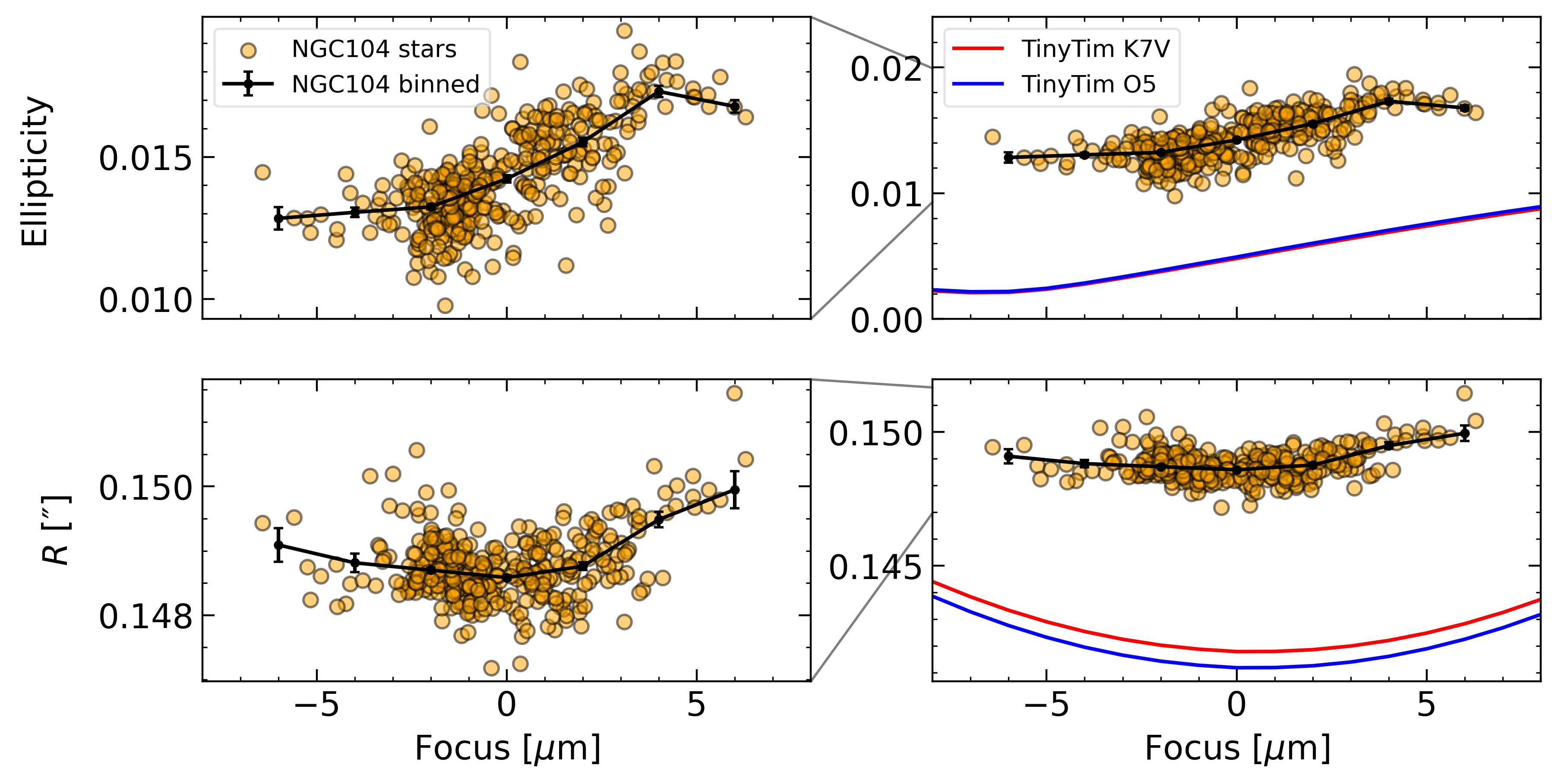}
    \caption{\textit{Left:} Variation of the average stellar ellipticity (top) and size (bottom) as a function of focus. Each orange circle represents a single F160W WFC3-IR NGC 104 frame. Median values for $2\mu m$-wide bins are displayed as black circles. In the range of $-5\mu m$ to $+5\mu m$, we find that the ellipticity and size of stars vary by 0.004 and $1\%$, respectively. \textit{Right:} A comparison of the ellipticity (top) and size (bottom) of observed stars and the TinyTim model (blue and red lines). The blue and red lines are TinyTim models for an O5-type star and K7V-type star, respectively. The TinyTim model is significantly rounder and smaller than observed stars suggesting that it is not an accurate representation of the PSF. 
    }
    \label{fig:tt_e_focus}
\end{figure*}
 
Figure \ref{fig:tt_e_focus} presents the median ellipticity (top panel) and size (bottom panel) of stars for each NGC 104 frame as a function of predicted focus (orange circles). The black circles and errorbars are the binned averages of the medians and the standard errors. We find that the median ellipticity of stars in the NGC 104 frames reaches a minimum at $-5\mu m$. For focus values that are lower, the statistics are lacking. For higher focus values, the ellipticity increases and the maximum change in ellipticity is 0.004 (from 0.013 to 0.017). In the bottom-left panel, we see that the dependence of size of the PSF is U-shaped with a minimum at $0\mu m$. From the lowest plateau to the highest peak, we estimate the increase in size to be $\mytilde1\%$.

As a step towards understanding the expected variations in the PSF caused by the focus, we generated PSF models with TinyTim \citep{2011krist}. The TinyTim software creates PSF models based on the theorized diffraction pattern of light through the optical components of the telescope. It allows the selection of detector, filter, stellar type, and image size. We selected the WFC3-IR detector, F160W filter, and an image size of 30 arcsec. PSFs of two stellar types, O5 and K7V, were created to investigate the difference in PSF properties for blue and red stellar types.  We generated PSFs with a range of focus values from $-8\mu m$ to $+8\mu m$ at intervals of 1$\mu m$. The output from TinyTim is a distorted PSF that has been convolved with the charge-diffusion kernel and sampled at 0\farcs13 per pixel. We placed the distorted TinyTim PSFs into blank FLT images (one for each focus) and drizzled them with the same parameters as the CANDELS and NGC 104 images. The TinyTim PSFs were extracted from the drizzled frames and their quadrupole moments were measured. 

The right panels of Figure \ref{fig:tt_e_focus} compare the ellipticity (top) and size (bottom) of NGC 104 stars to the drizzled TinyTim PSF models. The red and blue lines represent the TinyTim models for a K7V-type star and an O5-type star, respectively. In the top panel, the ellipticity of NGC 104 stars shows a similar trend to the TinyTim PSF ellipticity with increasing focus values leading to increasing ellipticity. However, the average ellipticity of the TinyTim model is 5-7 times lower than the ellipticity of stars from the NGC 104 images. The bottom panel compares the size \textit{R} of the TinyTim model with the median sizes of stars per frame. Both distributions are parabolic across the focus range investigated. The size of the TinyTim model is \mytilde 4\% smaller than the median size of stars. We suspect that the lack of a PRF in the TinyTim model is partly at fault for the difference in size.  

\subsection{SED shape} \label{sec:sed}
A major advantage of a space telescope is that it is not subject to the turbulent atmosphere of the earth and thus is diffraction limited. At the diffraction limit, the FWHM of an Airy disk is \begin{equation}
    \theta \approx 1.03\frac{\lambda}{D},
\end{equation} 
where $\lambda$ is the wavelength of light and D is the diameter of the aperture. Therefore, for a fixed aperture, the size of the PSF scales with the wavelength of light.

Astronomical imaging employs filters that allow a band of light to pass. For example, the HST/WFC3 F160W filter allows light from approximately 1.4$\mu$m to 1.7$\mu$m to reach the detector. At the extremities of this bandpass, the radius of the FWHM of the redder light is 1.2 times larger than the bluer light. The observed PSF is a superposition of the PSF of each wavelength of light that permeates the bandpass. In addition, the contribution of each wavelength is determined by the spectral energy distribution (SED) of the source and the wavelength-dependent sensitivity of the filter. Thus, the PSF is source dependent.

\begin{figure}
    \centering
    \includegraphics[width=0.5\textwidth]{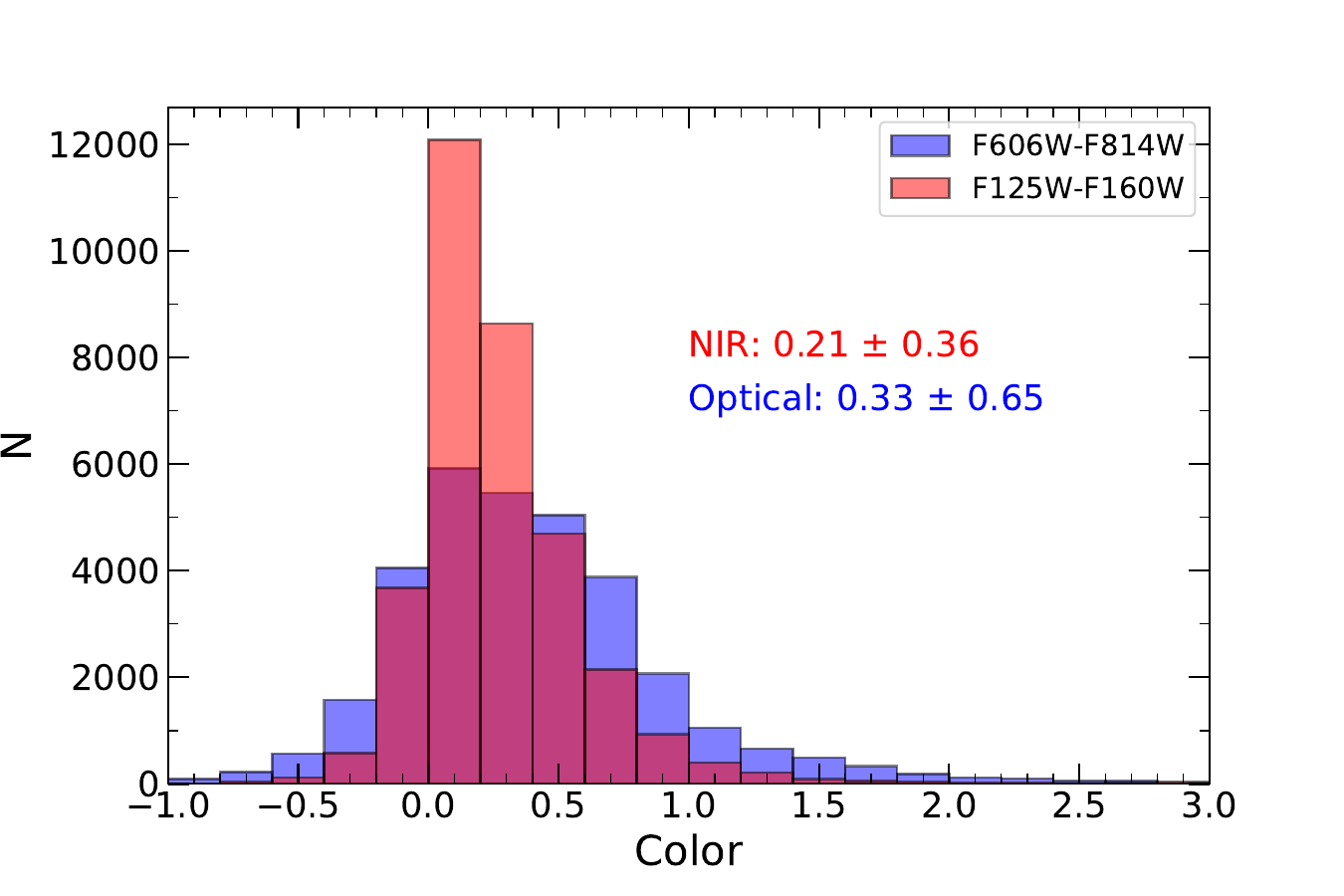}
    \caption{Comparison of galaxy colors in the optical (blue, F606W-F814W) to the NIR (red, F125W-F160W). The median and the standard deviation of the NIR and optical colors are $0.21\pm0.36$ and $0.33\pm0.65$, respectively. The lower variation of galaxy colors in the NIR is caused by less variation in galaxy SEDs in the bandpasses. This shows that the NIR PSF is less dependent on the SED of the galaxy than the optical PSF. }
    \label{fig:galaxy_color}
\end{figure}

The effect of the SED on the PSF is more important for telescopes that have wider bandpass filters. Furthermore, it affects filters that capture a region of the SED that has more variation in flux such as the rest-frame UV light that traces star formation. One way to demonstrate the severity that galaxy SEDs will have on the PSF is to analyze the variation in galaxy colors. Figure \ref{fig:galaxy_color} compares the optical (F606W-F814W) color of galaxies in the GOODS-S field with the NIR (F125W-F160W) color. The difference in scatter shows that galaxy SEDs in NIR have less variation than in optical. We found that all five CANDELS fields show similar distributions of NIR and optical colors. As a consequence, the PSF dependence on the wavelength is weaker in the NIR than in the optical.

In Figure \ref{fig:tt_e_focus}, we showed the difference in size and ellipticity for a star with an O5 and a K7V SED. We see an insignificant ($e<0.0005$) change in the ellipticity of the PSF for differing SED. When comparing the size of the blue (O5) PSF to the red (K7V) PSF, we find the redder PSF is 0.3\% bigger. We attempted to measure this effect in the NGC 104 images and were unable to statistically separate the sizes of blue and red stars. In part, the lack of statistics could be driven by low variety in the stellar types of the globular cluster NGC 104. In conclusion, assuming that the TinyTim model is a good approximation for the difference in PSF shape and size for differing stellar types, the bias caused by the SED is the smallest systematic effect in our study.

\subsection{Summary of Systematic Effects}
Table \ref{table:systematics} summarizes the ellipticity and size variations that we found for the WFC3-IR F160W PSF. In both ellipticity and size, undersampling causes the largest variations with a change in ellipticity and size of up to 0.02 and 3\%, respectively. The BFE modifies the size of stars at the 2\% level and the BRE modifies the ellipticity by 0.006. Telescope breathing leads to variations in the focus, which distort the shape of the PSF by 0.004 in ellipticity and 1\% in size. The least significant systematic effect in our analysis is variation in the PSF from the interplay of the filter throughput and the galaxy SED with variations in ellipticity and size expected to be at the 0.0005 and 0.3\% level, respectively.

\begin{table}[!h]
\caption {Relative Importance of Systematic Effects on PSF Parameters} \label{table:systematics}
\def\arraystretch{1.}
\begin{tabular}{ c  c  c }
\hline
\hline
Effect & E bias & Size bias           \\
\hline
Undersampling & 0.02     & 3\%             \\
BFE & 0.006    & 2\%            \\
Focus & 0.004    & 1\% \\
SED/Throughput & 0.0005    & 0.3\% \\
\hline

\end{tabular}

\end{table}

\section{WFC3-IR Point-Spread Function} \label{sec:PSF}
We have quantified the systematic effects that lead to spatial and temporal variations in the effective PSF. 
To proceed with a weak-lensing analysis, we aim to create PSF models that best represent all of the above effects. In this section, we present PSF models that are designed through a principle component analysis of archival imaging of NGC 104.

\subsection{Principal Component Analysis} \label{sec:PCA}
Modeling the PSF of HST has two issues associated with it. 
First, in typical extragalactic fields, there is an inadequate number of stars in a single observed frame due to the small field of view. Second, is the variation of the HST focus, and subsequently the shape of the PSF, during the 90 minute orbit.  Fortunately, one can rely on archival imaging of globular clusters to resolve both of these issues. Modeling the HST PSF from a principal component analysis (PCA) of stellar fields is a common technique \citep[see][]{2007jee,2009nakajima,2010schrabback}. Our method follows the same basic steps as the \cite{2007jee} technique, therefore we refer the reader to that work and subsequent descriptions of the technique on HST WFC3-IR for weak lensing \citep[][]{2017jee,2019kim,2020finner}. Here, we briefly summarize the steps that we took to model the F160W PSF with a PCA.

As mentioned in Sections \ref{sec:candels} and \ref{sec:ngc104}, we single drizzled the FLT frames for CANDELS and NGC 104 to the native detector orientation with the same drizzle parameters as the CANDELS mosaics. For each NGC 104 frame, SExtractor was run and a sample of stars was selected from the stellar locus in a size-magnitude plot. Stars near the saturation limit and stars that are too faint ($S/N<5$) were omitted from the sample. Postage stamp images (31x31 pixels) of each star were cut from the single-drizzled frame and their quadrupole moments were measured (Equation \ref{eq:quadrupole}). Stars with centroid (Equation \ref{eq:centroid}) differing by more than 0.3 pixels from the SExtractor measured position were removed because they tended to have a bright neighbor. This robust sample of stars was then stacked into an N star x 961 pixels array. Each star in the array was interpolated and shifted to be centered on the (15,15) coordinate pixel using a bicubic-spline interpolation. This interpolation is necessary because stacking uncentered stars would broaden the PSF. The star array was sigma clipped to remove $3\sigma$ outlying pixels and then the mean star was calculated and subtracted from the array to get the residual. At this point the important features that have been retained are the mean star, the residuals of each star, and the coordinates of each star.

A PCA was performed on the residual and the eigenvectors (principal components) that contain the 21 most significant signals were stored. The 21 principal components with the highest eigenvalues contain 98\% of the signal, with the remaining principal components dominated by noise. Utilizing the PCA result, the mean star, and the positions of the stars, PSF models can be generated that include the spatial dependence of the PSF. This is achieved by fitting the PCA result at the star positions with a third order polynomial, retrieving the coefficients, propagating the coefficients to generate the spatial dependence of the PSF model at the desired (x,y) coordinates, and then adding it to the mean PSF. Following this recipe, we can produce a 31 pixel x 31 pixel PSF model at any x,y coordinate. 

A key step is to find the NGC 104 PCA result that best matches the observed CANDELS frame in order to properly account for the temporal variation of the PSF. For the ACS, \cite{2007jee} suggest that the NGC 104 frame that best fits a science frame can be found by minimizing the quadrupole moments as follows:

\begin{equation}\label{eq:jee_min}
    \chi^2 = \sum_{ij}{\frac{\left(Q_{ij} - Q'_{ij}\right)^2}{\sigma_{ij}^2}}
\end{equation}
where the primed values are from the PSF model and the un-primed values are measurements from stars in the CANDELS frames. To estimate the noise ($\sigma_{ij}$) for each CANDELS star, we performed 100 measurements of the quadrupole moments with Gaussian noise of the rms level applied. To apply Equation \ref{eq:jee_min}, stars that were selected from the mosaic images, see Appendix \ref{app:star_selection}, were found in the single-drizzled CANDELS frames and their quadrupole shapes were measured. Using their coordinates, we derived the spatially-correct PSF model for all possible NGC 104 templates and selected the template with the lowest $\chi^2$. Once the best-fit frame was found, we generated PSF models for each galaxy in the frame. At this stage, we have one PSF model for each galaxy in each CANDELS frame. Therefore, an additional step was taken to stack the component PSFs of each frame into coadded PSFs that are usable for the coadded mosaics. The final PSFs thus include the spatial and temporal variation and have been stacked into coadded PSFs.

    
\begin{figure*}[!ht]
    \centering
    \includegraphics[width=\textwidth]{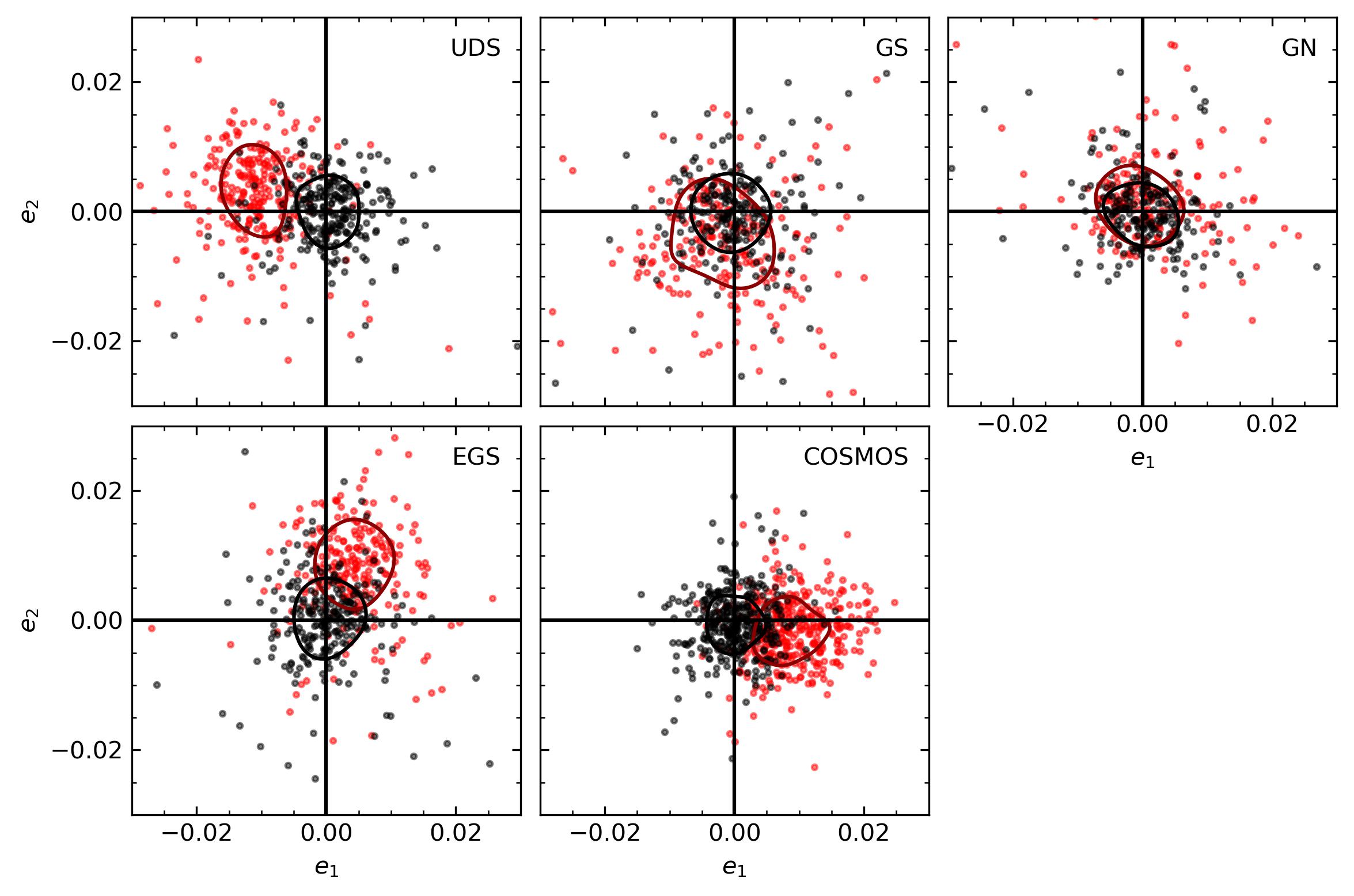}
    \caption{Ellipticity correction for stars in the five CANDELS fields. The red circles are the measured ellipticity of stars from each coadded mosaic. The black circles are the residual ellipticities after subtracting the measured ellipticity of the PSF models from their respective stars (red circles). For all fields, the PSF model corrects for the average ellipticity (shifts the centroid of the distribution toward zero) and corrects for the spatial variation of the ellipticity (reduces the scatter)}. The residual $e_1$ and $e_2$ components are summarized in Table \ref{tab:psf_corrections} and on average they leave residuals of $|de_1|<0.0005\pm0.0003$ and $|de_2|<0.0005\pm0.0003$, which is sufficient for our future weak-lensing study of massive overdensities.
    \label{fig:res_ellipticity}
\end{figure*}
\section{PSF Model Validation} \label{sec:validation}
Correcting for the PSF is a critical step in weak lensing. In the previous sections, we discussed the systematic effects that should be included in an effective PSF and then modeled the effective PSFs of the five CANDELS fields. In this section, the validity of the PSF models will be inspected by comparison with stars in the mosaic images. 


\subsection{PSF Ellipticity Tests}
The observations of each of the five fields were obtained in a variety of observing modes and plans. For example, the UDS field was primarily observed in two visits with the telescope roll angle close to 90 degrees and a two-point dither pattern in each visit. 
This observation plan leads to a stacked PSF that elongates in the north-south direction. In other fields, such as GOODS-S, the roll angle between the observation epochs is different, which lead to a rounder PSF when coadded.

The two properties of the PSF that are of major concern to our future weak-lensing analysis are the ellipticity and size. A perfect model will fully encapsulate all of the spatial and temporal variations of the PSF and be able to remove their effect from galaxy shape measurements. For insight into how well our PSF models have captured these variations, we calculate the residual ellipticity and size by subtracting the measurements of stars in the coadded mosaic from the measurements of their corresponding PSF models.

Figure \ref{fig:res_ellipticity} shows the measured ellipticity of stars (red circles) and the residual ellipticity (black circles) after subtracting the ellipticity of the PSF model. The median values and their $1\sigma$ uncertainties are presented in Table \ref{tab:psf_corrections}. The UDS was imaged with a 90 roll angle. Thus, the PSF is dominated by a negative $e_1$ component. COSMOS was imaged at approximately 0 or 180 degrees roll angles, which imparts a postive $e_1$ component to the PSF. For EGS, the imaging was completed with roll angles near 45 and 225 degrees. In this case, the PSF shows a significant $e_2$ component.  Finally, for GOODS-S and GOODS-N, the roll angles varied with each epoch and the result is a rounder PSF. The three fields with observations that were designed with nearly constant roll angles (UDS, EGS, and COSMOS)  
show that the PSF model corrects for a large ellipticity component. For GOODS-S and GOODS-N, the PSFs make a smaller correction to the ellipticity. In all five fields, the PSF models lower the scatter in the ellipticity. Overall, our PSF models are able to account for the ellipticity of the PSF in the five fields to $|e_1|<0.0005\pm0.0003$ and $|e_2|<0.0005\pm0.0003$. 

\begin{table*}[!ht]
\caption {PSF Ellipticity Correction} \label{table:psf_corrections}
\def\arraystretch{1.}
\centering
\begin{tabular}{ l  r  r  r  r}
\hline
\hline
\multicolumn{1}{l}{ } &
\multicolumn{2}{c}{Before correction} &
\multicolumn{2}{c}{After correction}\tabularnewline
Field & $e_1$ [$10^{-4}$] & $e_2$ [$10^{-4}$] & $e_1$ [$10^{-4}$] & $e_2$ [$10^{-4}$]           \\
\hline
UDS & $-108.8^{+3.2}_{-3.2}$     & $30.2^{+3.5}_{-4.0}$ & $4.8^{+2.9}_{-2.9}$ & $1.5^{+2.7}_{-3.1}$ \\
GOODS-S & $-13.9^{+5.3}_{-5.3}$     & $-44.8^{+4.9}_{-5.6}$ & $-4.0^{+3.9}_{-4.1}$ & $-1.7^{+3.4}_{-4.2}$ \\
GOODS-N & $3.8^{+5.9}_{-3.6}$     & $12.9^{+3.8}_{-3.8}$ & $-0.8^{+3.7}_{-3.6}$ & $-4.7^{+3.3}_{-3.6}$ \\
EGS & $45.4^{+3.7}_{-3.4}$     & $82.6^{+3.8}_{-4.4}$ & $0.3^{+3.5}_{-2.9}$ & $3.0^{+3.6}_{-3.8}$ \\
COSMOS & $83.8^{+3.5}_{-2.4}$     & $-13.8^{+2.3}_{-2.6}$ & $-1.2^{+2.6}_{-2.1}$ & $-4.8^{+2.3}_{-2.2}$ \\
\hline

\end{tabular}
\label{tab:psf_corrections}
\end{table*}

\begin{figure*}
    \centering
    \includegraphics[width=\textwidth]{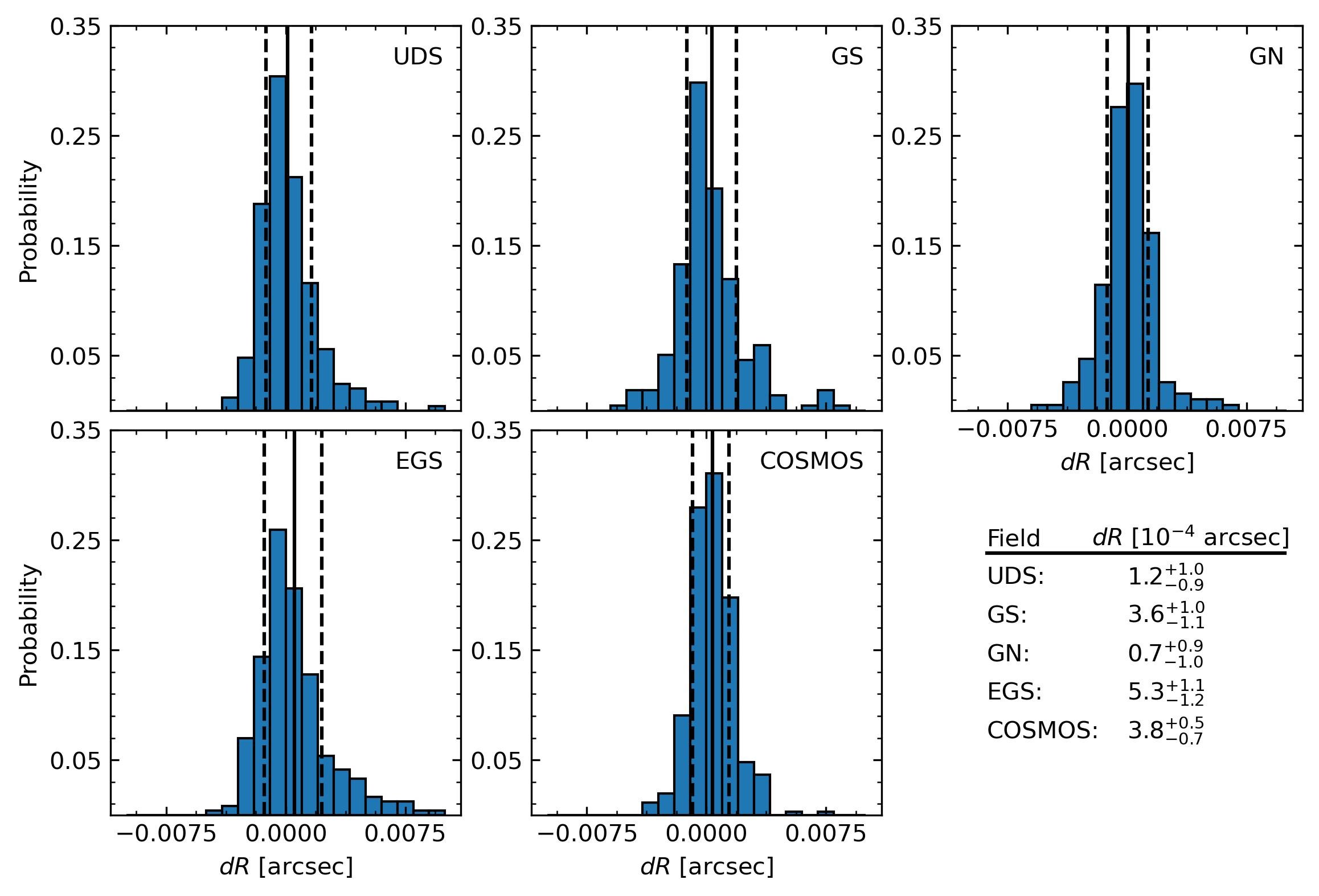}
    \caption{Residual PSF size after subtracting the size of the PSF models from that of the stars. The PSF models match the size of stars to $\lesssim 0.3\%$. This residual in size is below the dominant systematic effects of undersampling and the BFE and thus is acceptable for our future weak-lensing analysis.}
    \label{fig:res_size}
\end{figure*}

Figure \ref{fig:res_size} displays the size residuals after subtracting the PSF models from the stars. The black vertical line is the median value and the vertical dashed lines are the $1\sigma$ uncertainties. The median residual sizes are $dR < 0.0005\pm0.0001$ for all five fields. This residual will introduce a small multiplicative shear bias into weak-lensing shape measurements that will be calibrated through image simulations.  

\section{Summary} \label{sec:discussion}
Weak lensing in the NIR benefits from an increased number density of galaxies and lowered uncertainty on shape measurements. Performing weak lensing in the NIR enables the study of overdensities to higher redshift, with high precision. To date, no NIR weak-lensing studies in blank fields have been achieved. However, before the end of the decade, Euclid will image 15,000 square degrees of the sky in the NIR and the Nancy Grace Roman Telescope will constrain cosmological parameters with NIR weak lensing. Thus, we have initiated a NIR weak-lensing (NIRWL) analysis of the five CANDELS fields (UDS, GOODS-S, GOODS-N, EGS, and COSMOS) to identify systematics that may limit future weak-lensing surveys.

Comparing the Treasury CANDELS images to the Gaia catalog, we determined that the astrometry was poor with the average offset being as large as $0\farcs26$ from Gaia star positions. Furthermore, some of the fields show a systematic variation of the offset. Utilizing the positions and proper motions of stars in the Gaia catalog, we updated the alignment of the five CANDELS fields. The newly aligned CANDELS images have an absolute astrometric accuracy of $\lesssim0.02\pm0.02\arcsec$ relative to the Gaia catalog.

We investigated the systematic effects of the WFC3-IR detector and PSF. 

\begin{itemize}
    \item The PSF for HST WFC3-IR filters is undersampled. We detected the sub-pixel dependence of the PSF shape in F160W-filter imaging of NGC 104. By isolating the PSF centroid in FLT images, we were able to show that sub-pixel \textit{x} and \textit{y} shifts of the centroid from the center of a pixel lead to variations in the PSF ellipticity and size by 0.02 and $3\%$, respectively. This is the largest systematic variation that we found in our study. We simulated a realistic weak-lensing technique for shape fitting with undersampled images and found that sub-pixel shifts of a galaxy lead to variations in the multiplicative galaxy shape bias that depend on galaxy centroid, galaxy size, galaxy shape, sampling rate, PSF size, and PSF shape. Simulating the measured properties of the UDS field, we estimate the multiplicative shear bias of the UDS to be -0.025. We find that limiting the galaxies to those with half-light radius above $0\farcs15$ can reduced the multiplicative shear bias to approximately 0.002.
    
    \item  Analyzing the size of stars in the WFC3-IR NGC 104 images showed that the brighter-fatter effect causes the size of the PSF to increase by 2\% in the 22nd to 17th magnitude range. Along with the brighter-fatter effect, we detected a decrease in the ellipticity of stars by 0.006 over the same magnitude range. We call this the brighter-rounder effect.  
    
    \item Focus variations that are caused by the orbit of the HST lead to temporal changes in the shape of the PSF. Employing the focus predictions from the HST website, we found that size and ellipticity of stars in F160W vary by 1\% and 0.004, respectively.
    
    \item Chromatic dependence of the PSF arises from the interplay of chromatic diffraction, galaxy SEDs, and filter throughput. We showed that the chromatic dependence of the NIR-bands is lower than in optical because the scatter of galaxy colors in the NIR bands is less than in optical. As a second test, we utilized the TinyTim PSF modeling code to show that variation in the PSF size and ellipticity based on blue (05) and red (K7V) stellar types changed the ellipticity and size by 0.0005 and 0.3\%, respectively. Therefore, chromatic variation of the NIR PSF is a subdominant effect.
    
\end{itemize}

We created PSF models for the F160W imaging of the five CANDELS fields. As with most HST observations, the CANDELS fields lack sufficient numbers of stars to properly model the PSF. Therefore, as others have done, we utilized archival imaging of the globular cluster NGC 104 for PSF modeling. To curtail the brighter-fatter and brighter-rounder effects, we selected stars well below the saturation limit for PSF modeling. The temporal aspect of the PSF was modeled by finding the best-fit NGC 104 frame through a comparison of the quadrupole moments of stars. We validate our PSF models by subtracting the measured properties of the PSF models from the measured properties of their star counterparts in the CANDELS fields. The average residual ellipticities are $|e_1|<0.0005\pm0.0003$ and $|e_2|<0.0005\pm0.0003$ and the average residual size is $|dR|<0.0005\pm0.0001$. These residual are acceptable for our future NIRWL weak-lensing analysis.






\begin{acknowledgments}
This work was supported by the National Research Foundation of Korea(NRF) grant funded by the Korea government(MSIT), NRF-2022R1C1C1008695. This work has made use of data from the European Space Agency (ESA) mission
{\it Gaia} (\url{https://www.cosmos.esa.int/gaia}), processed by the {\it Gaia}
Data Processing and Analysis Consortium (DPAC,
\url{https://www.cosmos.esa.int/web/gaia/dpac/consortium}). Funding for the DPAC
has been provided by national institutions, in particular the institutions
participating in the {\it Gaia} Multilateral Agreement. 

For the purpose of open access, the authors have applied a Creative Commons Attribution (CC BY) licence to any Author Accepted Manuscript version arising from this submission.

\end{acknowledgments}

\vspace{5mm}
\facilities{HST(WFC3)}

\software{Astropy \citep{2022astropy},  
          SExtractor \citep{1996bertin},
          Drizzlepac \citep{2010fruchter, 2012gonzaga}}

\appendix

\section{CANDELS mosaics}\label{app:candels_redo}
Weak lensing relies on well-aligned and distortion-free images. Misalignments and distortions can lead to unwanted shear signals caused by the rounding or elongation of objects. Comparing the proper-motion-corrected Gaia star positions to the star positions in the CANDELS Treasury mosaics reveal significant astrometric offsets. These astrometric offsets are likely caused by the lack of a high-precision reference catalog at the time the mosaics were created. 

\begin{figure}
    \centering
    \includegraphics[width=0.7\textwidth]{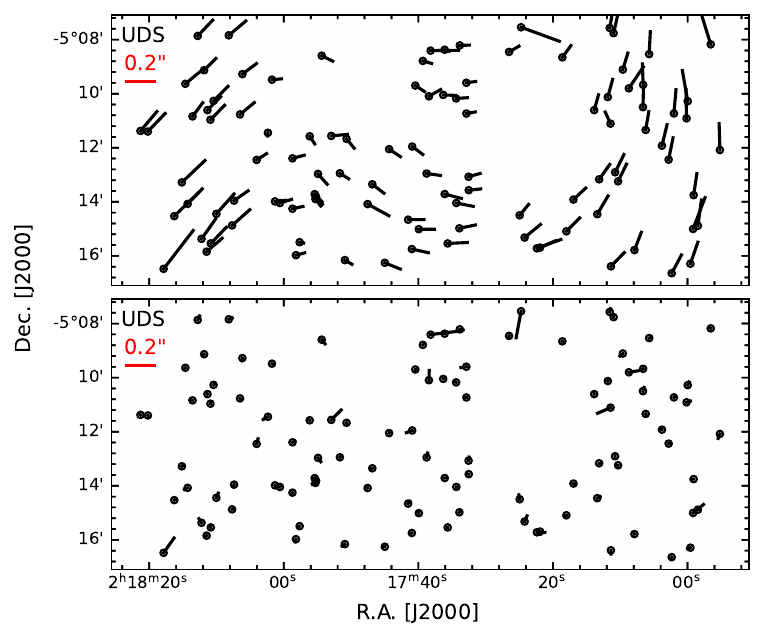}\\
    \caption{Astrometric offset of stars in the Gaia catalog to stars in the mosaics from CANDELS Treasury (top panel) and our work (bottom panel). The stars have been corrected for proper motion. The CANDELS Treasury positions show a systematic offset that varies from \mytilde0\farcs1 in the center of the image to \mytilde0\farcs25 on the outskirts. This systematic variation has been corrected in our mosaic. On average, the astrometry of the UDS field is improved from 0\farcs13 to 0\farcs01.}
    \label{fig:uds_alignment}
\end{figure}

\begin{figure}
    \centering
    \includegraphics[width=0.9\textwidth]{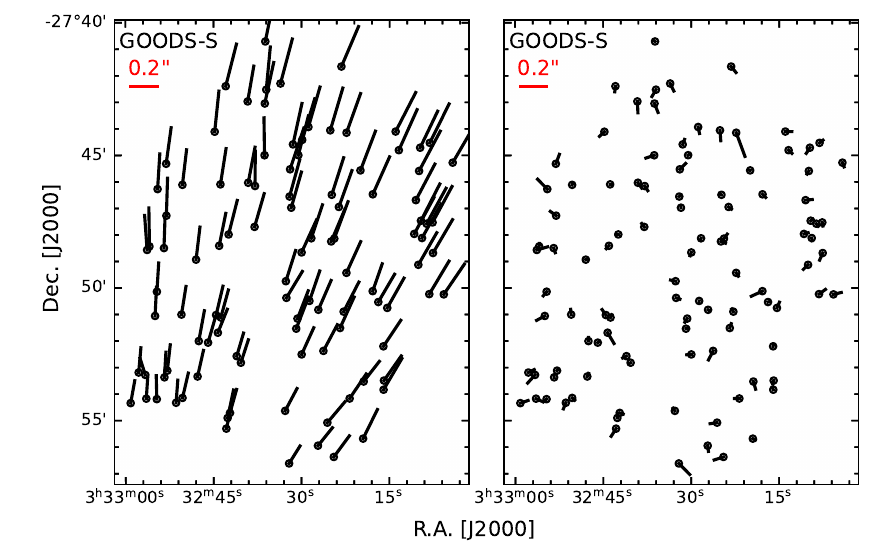}
    \caption{Astrometric offset of stars in the Gaia catalog to stars in the mosaics from CANDELS Treasury (left panel) and our work (right panel). The CANDELS Treasury positions show a nearly constant offset from the Gaia positions by \mytilde0\farcs26. Our mosaic has removed the offset. On average, the astrometry of the GOODS-S field is improved from 0\farcs26 to 0\farcs02.}
    \label{fig:gs_alignment}
\end{figure}

\begin{figure}
    \centering
    \includegraphics[width=0.9\textwidth]{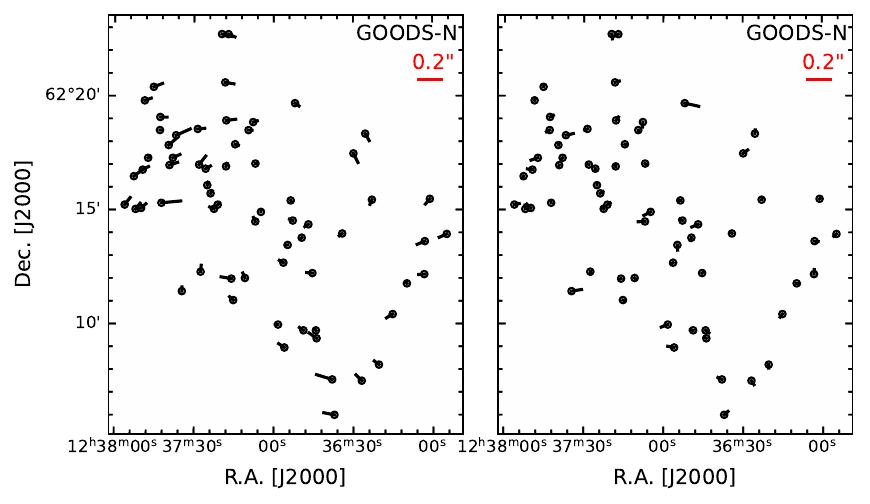}
    \caption{Astrometric offset of stars in the Gaia catalog to stars in the mosaics from CANDELS Treasury (left panel) and our work (right panel). The Treasury mosaic is well aligned but does show a slight systematic variation at the edge of the mosaic. Our mosaic has removed the systematic variation and, on average, the astrometry of the GOODS-N field is improved from 0\farcs05 to 0\farcs02.}
    \label{fig:gn_alignment}
\end{figure}

\begin{figure}
    \centering
    \includegraphics[width=\textwidth]{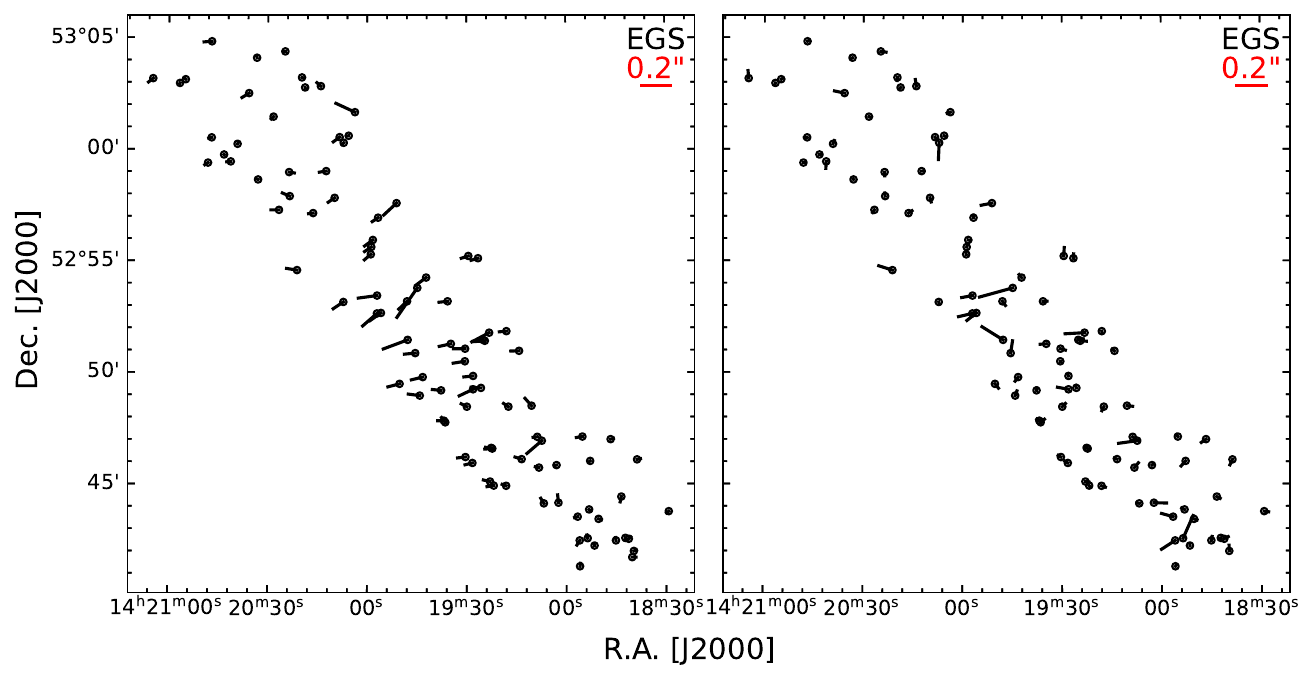}
    \caption{Astrometric offset of stars in the Gaia catalog to stars in the mosaics from CANDELS Treasury (left panel) and our work (right panel). The Treasury mosaic for the EGS is well aligned but stars near the center of the image have larger offsets than stars near the boundary. Our new alignment to Gaia has corrected this variation. On average, the astrometry of the EGS field is improved from 0\farcs07 to 0\farcs03.}
    \label{fig:egs_alignment}
\end{figure}

\begin{figure}
    \centering
    \includegraphics[width=0.7\textwidth]{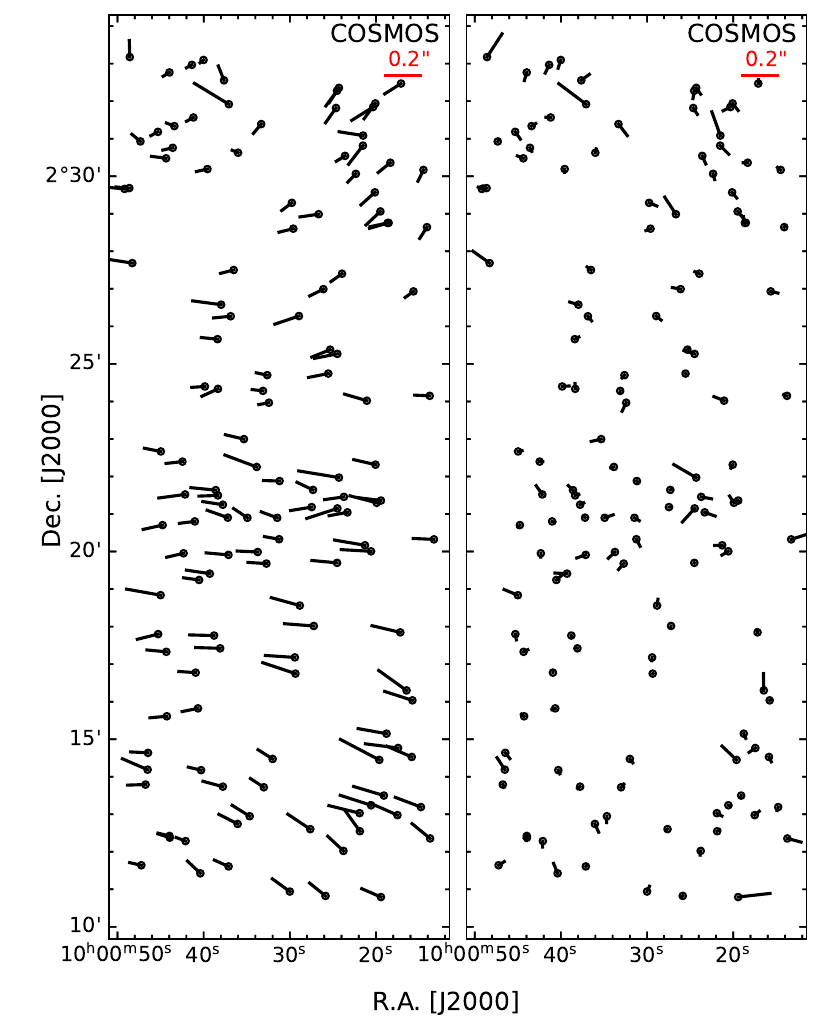}
    \caption{Astrometric offset of stars in the Gaia catalog to stars in the mosaics from CANDELS Treasury (left panel) and our work (right panel). The Treasury mosaic of COSMOS has a \mytilde0\farcs11 shift from the Gaia positions. Our mosaic has removed the shift. The average astrometry of the COSMOS field is improved from 0\farcs11 to 0\farcs02.}
    \label{fig:cosmos_alignment}
\end{figure}

The astrometric offsets that we detect are systematic. In the UDS mosaic (left panel of Figure \ref{fig:uds_alignment}), a clear variation is detected with stars on the edges of the mosaic having larger offset from the Gaia positions. This type of variation can be caused by an image registration strategy that aligns adjacent images in radially growing steps. Similar patterns are also found in the GOODS-N and COSMOS mosaics (left panels of Figures \ref{fig:gn_alignment} and \ref{fig:cosmos_alignment}), but to a lesser extent. Bulk translational offsets are also found. GOODS-S (left panel of \ref{fig:gs_alignment}) is the prime example of a large translational offset. This type of offset could be caused by aligning a mosaic to a single star or a poor reference catalog.  The right panels of Figures \ref{fig:uds_alignment} - \ref{fig:cosmos_alignment} show the positions are stars in our mosaics with astrometry aligned to the proper-motion-corrected Gaia catalog. The systematic variations in absolute astrometry that are present in the Treasury mosaics have been improved with no coherent pattern found in the new mosaics.

\section{Star Selection}\label{app:star_selection}

\begin{figure}
    \centering
    \includegraphics[width=\textwidth]{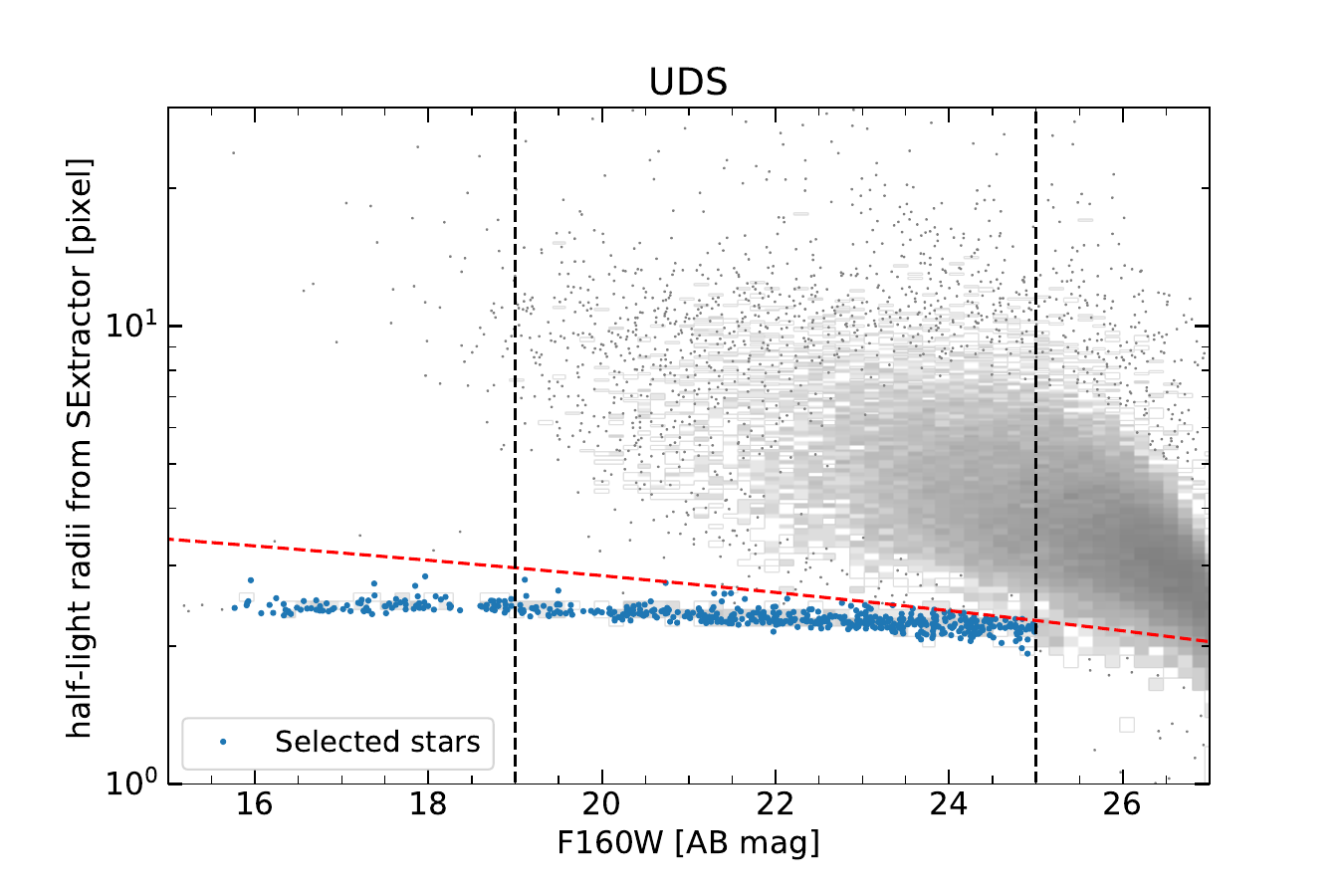}
    \caption{Half-light radius [pixel] vs. F160W magnitude diagram of sources in the CANDELS/UDS field. Darker grey represents a denser region of sources in this diagram. There is a tight sequence of stars which is mostly below the red line defined as half-light radius $< -0.115\times F160W + 5.15$. The additional magnitude cuts are vertical dashed black lines, F160W = 19 AB (Saturation limit) and 25 AB magnitudes, respectively. Blue points indicate stars classified from our star selection criteria.}
    \label{fig:candels_stars_uds}
\end{figure}

Stars tend to follow a stellar locus in a size-magnitude diagram because they are point sources and thus take on the size of the PSF \citep{Barro2019, Skelton2014}. We have identified stars using SExtractor half-light radius (FLUX\_RADIUS) and F160W MAG$\_$AUTO obtained from publicly available CANDELS F160W photometric catalogs of five CANDELS fields. Figure~\ref{fig:candels_stars_uds} compares the SExtractor half-light radius against the F160W magnitude for all sources in the CANDELS/UDS catalog. One can clearly identify a distinct stellar locus in this diagram. For robust star selection, we restrict the sample to 
\begin{enumerate}
    \item those located below a cut (red line in Figure~\ref{fig:candels_stars_uds}) adopted from CANDELS/GOODS-N \cite{Barro2019}, half-light radii $< -0.115\times F160W+5.15$ to all five fields 
    \item have a SExtractor FLAG = 0 to avoid bad photometry
    \item have F160W $<$ 25 AB mag 
    \item exclude AGNs with AGN flags obtained from the CANDELS catalog 
    \item consider an object having spectroscopic redshift = 0 as a star
    \item visually exclude sources blended with nearby objects.
\end{enumerate}
The blue points in Figure~\ref{fig:candels_stars_uds} are the stars selected from the above restrictions. They are clearly separated from galaxies down to F160W$\sim$25 mag. Very bright suffer saturate the detector. By cross-correlating the surface brightness profile of each star with the median surface brightness profile of stars, we were able to detect the saturation limit. We found that stars with F160W $<$ 18.5 AB magnitudes are saturated in WFC3. Thus, in PSF modeling and astrometric calculations, we only use stars having 18.5 $<$ F160W $<$ 25 AB magnitudes to avoid saturated stars. This star selection criteria was applied to all five fields. 

\section{Further Undersampling Examples}\label{app:undersampling}
Here, we present two additional figures that provide insight into undersampling in the WFC3-IR detector. 

\begin{figure*}
    \centering
    \includegraphics[width=\textwidth]{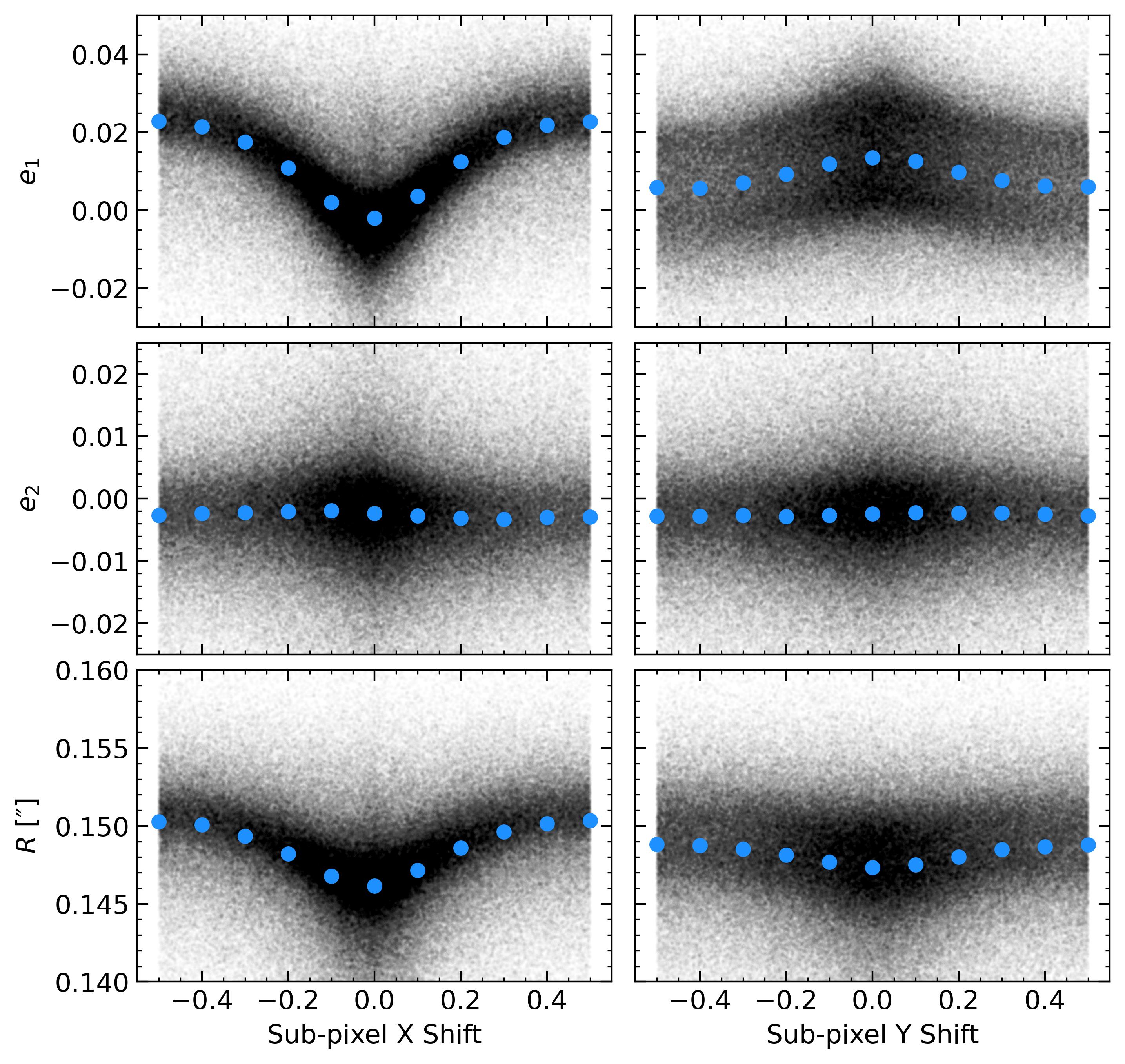}
    \caption{PSF properties of drizzled frames from observations of NGC 104. The intrinsic ellipticity of the PSF is in the detector \textit{x} direction. \textit{Top:} Sub-pixel shifts in x lead to increased ellipticity of the PSF and sub-pixel shifts in y decrease the ellipticity. \textit{Middle:} Sub-pixel shifts have minimal effect on the $e_2$ component. \textit{Bottom:} Shifts in either direction lead to an increase in size of the PSF.}
    \label{fig:undersampling_driz}
\end{figure*}

Figure \ref{fig:undersampling_driz} shows the quadrupole measurements of stars in NGC 104 after removing geometric distortion by drizzling. The sub-pixel centroid positions have been converted from the drizzled position back to the position in the FLT frame because that is where undersampling occurs. In the top-left panel, the $e_1$ component is positive, which signifies an elongated PSF along the \textit{x} direction. When the centroid is displaced in the \textit{x} (\textit{y}) direction, $e_1$ increases (decreases) as expected for an image with an intrinsic \textit{x}-direction elongation. The $e_2$ component has no noticeable change in shape with centroid displacements. The bottom row shows that size increases for both displacements.

\begin{figure*}
    \centering
    \includegraphics[width=\textwidth]{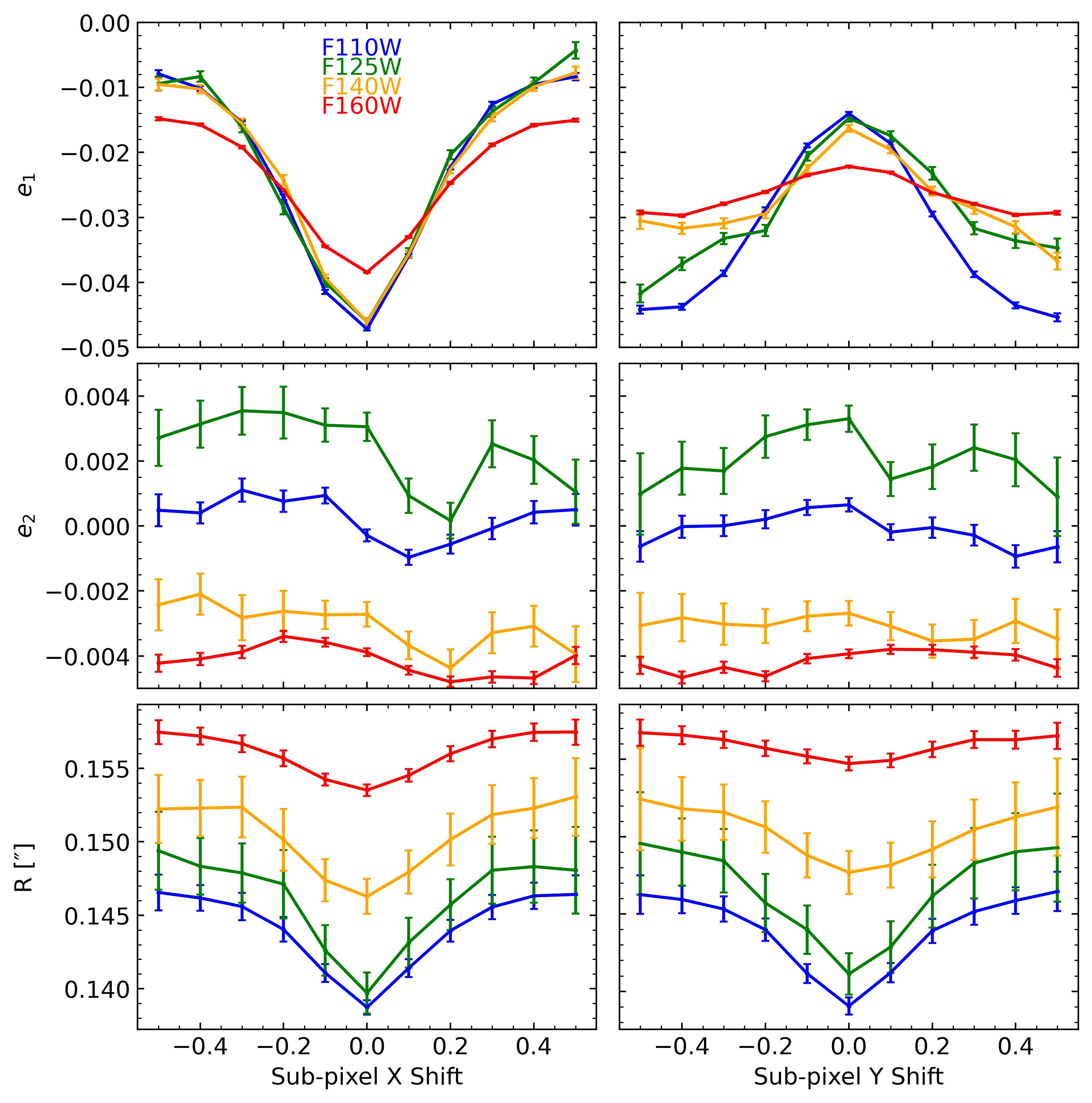}
    \caption{PSF properties of FLT frames from observations of NGC 104. Bluer filters are more subject to the undersampling effect because their PSF is smaller. Clearly, the F160W filter is the best choice for mitigating the undersampling effect in WFC3-IR imaging. Future missions should consider the impact that undersampling could have on the science when selecting filter and plate scales.}
    \label{fig:undersampling_flt_bycolor}
\end{figure*}

In Figure \ref{fig:undersampling_flt_bycolor}, we present the undersampling effect for four HST filters: F105W (blue), F125W (green), F140W (orange), and F160W (red). As expected, the bluer the filter the more it is subject to undersampling. 

\clearpage

\bibliography{sample631}{}
\bibliographystyle{aasjournal}








\end{document}